\documentclass[twocolumn,showpacs,preprintnumbers,amsmath,amssymb]{revtex4}

\usepackage{graphicx}
\usepackage{dcolumn}
\usepackage{bm}

\begin{document}


\title{GLHUA Electromagnetic Invisible Double Layer Cloak \\
With Relative Parameters Not Less Than 1 
and GL No Scattering Inversion \\
Radial coordinate $R$ in sphere coordinate can be negative in a negative space world}

\author{Jianhua Li}
 \altaffiliation[Also at ]{GL Geophysical Laboratory, USA, glhua@glgeo.com}
\author{  Feng Xie, Lee Xie, Ganquan Xie}%
\affiliation{%
GL Geophysical Laboratory, USA
}%

\hfill\break
\author{Ganquan Xie}
\affiliation{
Chinese Dayuling Supercomputational Sciences Center, China \\
Hunan Super Computational Sciences Society, China
}%

\date{December 31, 2016}

\begin{abstract}
We have know the radial coordinate $R$ is non negative in the polar coordinate since our middle school,the radial coordinate  $R$ is non negative in the sphere coordinate since our high school, and  since our undergraduate university.  Up to now, in all publications of EM, acoustic, and seismic wave modeling and inversion, the radial coordinate $R$ is non negative. In this paper and my paper arXiv:1706.10147, first and surprise in the world,  we propose that the radial coordinate $R$ can be negative in 2D polar coordinate, in 3D and ND sphere coordinate in EM and Acoustic and Seismic Wave propagation modeling and Inversion. 
 we discover that the radial variable in the sphere coordinate can be negative. The space in which the radial variable in the sphere coordinate is positive or zero is called positive space, that is real three dimensional space we are living.  The space in which the radial variable in the sphere coordinate is  negative is called negative space. The positive space is visible, but  the negative space is invisible.  In the acoustic equation in the sphere coordinate [32], radial electric wave equation and radial magnetic wave equation in the spherical coordinate [1] [31], if the radial variable $R$ is  changed to $-R$ , the equations are invariant, if the acoustic wave field solution is $u(R,\theta ,\phi )$ , the $u(-R,\theta ,\phi )$  is also solution of the homogeneous acoustic equation. Similar,  if radial electric field $E_r (R,\theta ,\phi )$  and radial magnetic wave $H_r (R,\theta ,\phi )$  are solution of the radial electric and radial magnetic equations, then 
 $E_r (-R,\theta ,\phi )$
and $H_r (-R,\theta ,\phi )$
  are also the solution of the homogeneous radial electric and radial magnetic equations  respectively, Therefore the acoustic equation and radial electric and magnetic equations and their solutions  in the sphere coordinate in the positive space with $R \ge 0$  can be analytical continuation into the negative space with $R < 0$ .     
We give a definition that the world of the non negative spherical radial coordinate  $R$ is the positive world ("Yang Jian" in China), inversely, the world of the negative spherical radial coordinate $R$ is the negative world ("Ying Jian" in China).  The door between the positive and negative is the zero $0$. In the normal sciences, the door $0$ is closed. We proposed the radial 
transform GLHUANPI in this paper and GLHUANPII transformation in my paper 
arXiv:1706.10147 [33] to open the door and translate the negative space to the positive space. The detailed contents of the negative space is proposed in my
paper arXiv:1706.10147 [33] 
No scattering is the zero scattering. Using GILD and GL no scattering modeling and inversion method, we find a class of the nonzero solution of the zero scattering nonlinear inversion equation and use it to create our GLHUA cloak. The nonzero solution of the zero scattering nonlinear inversion equation form more complicated infinite class than the nonzero solution of the linear operator equation. GLHUA cloak and GLLH cloak are different class of the nonzero solution of the zero scattering electromagnetic inversion, both is different from Pendry cloak. This paper is different from paper arXiv1005.3999 and is different from other cloak papers. In current general science, the visible natural science and invisible thinking and social science are main object. Inversely, the invisible natural science and visible thinking and social science are main object in a new novel super science. Our GLHUA practicable double layer cloak and Easton LaChappelles mind control robot hands show that the new novel super science is being born.  
In this paper, we discovered a class of GLHUA electromagnetic invisible double layer cloak with relative electromagnetic parameter not less than 1 for each layer, and each layer with any thickness. The major new ingredients in this paper are: (1) In the outer layer cloak 
$ {R_1}  \le r \le {R_2} $
the relative radial electric permittivity and radial magnetic permeability equal to 1, the relative angular electric permittivity and magnetic permeability are   $
\varepsilon _\theta   = \varepsilon _\phi   = \mu _\theta   = \mu _\phi   = \frac{1}{2}\left( {\left( {\frac{{r - R_1 }}{{R_2  - R_1 }}} \right)^\alpha   + \left( {\frac{{R_2  - R_1 }}{{r - R_1 }}} \right)^\alpha  } \right)
$, 
the relative parameters and their derivative are continuous across the outer boundary, 
$r = R_2 $ of the outer layer cloak; (2) in the inner layer cloak, 
$R_0  \le r \le R_1 $, the relative radial electric permittivity and radial magnetic permeability equal to 1, the relative angular electric permittivity and magnetic permeability are   
$
\varepsilon _\theta   = \varepsilon _\phi   = \mu _\theta   = \mu _\phi   = \frac{1}{2}\left( {\left( {\frac{{R_1  - r}}{{R_1  - R_0 }}\frac{{R_0 }}{r}} \right)^\alpha   + \left( {\frac{{R_1  - R_0 }}{{R_1  - r}}\frac{r}{{R_0 }}} \right)^\alpha  } \right)
$,
the relative parameters and their derivative are continuous across the inner boundary,
$r = R_0 $
, of the inner layer cloak; (3) GLHUA cloak is created by our Global and Local (GL) full electromagnetic wave modeling and no scattering inversion. The new idea and method of creating GLHUA double invisible layer cloak is different from transform optical method; (4) GL electromagnetic filed concept and GL electromagnetic wave field equation and full GL Green's function equation and full GL integral equation and angular Green's function equation and angular Greens function matrix are proposed; (5) With weight
$
\sin \theta 
$, the spherical surface weighted integral of GL electromagnetic wave field equal to zero; (6) We proposed GLHUA pre-cloak material conditions (6.1)-(6.4) and rigorously proved that electromagnetic wave field is going to zero when r is going to zero in invisible virtual sphere. We rigorous proved that in outer layer cloak $\left[ {R_1 ,R_2 } \right]$, when r decreasingly going to boundary
$r = R_1 $, the electromagnetic wave field is going to zero. The incident electromagnetic wave excited in outside of GLHUA cloak can not be disturbed by cloak, and the electromagnetic wave can not propagation penetrate into the concealment; (7) Reciprocally, in inner layer cloak 
$\left[ {R_0 ,R_1 } \right]$
, when r is increasingly going to boundary 
$r = R_1 $, incident electromagnetic wave excited in inside of the concealment is going to zero, the electromagnetic wave propagation can not arrive to the boundary $r = R_1 $, and the wave can not propagation to outside of the cloak; the incident electromagnetic wave excited in inside of the concealment can no be disturbed by cloak; (8) Theoretical proof and many simulations by GL full wave no scattering and inversion show that GLHUA electromagnetic double layer cloak is a practicable complete invisible cloak without exceeding light speed propagation and with relative parameters not less than 1 and reciprocal principle is satisfied which is totally different from Pendry cloak and other cloaks; (9) The strange double layer cloak phenomena in Figure 1 and Figure 2 discovered in 2000 are image of the nonzero solution of GILD no scattering inversion which cloaked the object of GILD scattering inversion. After 16 year hard research works, in this paper, GLHUA invisible electromagnetic double layer cloak presented a novel model and theory for explaining the double layer cloak phenomena discovered in 2000 [3]; (10) We created GILD and GL no scattering modeling and inversion method to discover, create and prove new cloak with relative parameter not less than 1 and with phase velocity less than light speed and tends to zero at the boundary $r = R_1 $. The idea and no scattering modeling and inversion method to create GLHUA cloak is novel and different from all other cloak publications by other researchers. In my paper arXiv:1706.10147 [33], we found  an exact full EM wave propagation solution of the EM equation with the practicable GLHUA double layer perfect invisible (stealth) cloak , which is great breakthrough progress. The exact solution of EM equation with GLHUA and GLLH double layer cloak is indisputable evidence to prove that our GLHUA GLLH double layer invisible cloak model is completely  practicable perfect invisible cloak physics mathematical model. A novel exact EM wave propagation in the GLLH and GLHUA cloak is published in my paper [33] and brief report here.
It is totally different from the transformation optics, and completely do not use transformation optics, by using GL no scattering modeling and inversion [1][31], in this paper and my paper [33], we discover a new GLHUA outer annular layer EM invisible cloak with relative parameter not less than 1. In GLHUA outer layer invisible cloak , $R_1  \le r \le R_2 $, the relative radial electric permittivity and magnetic permeability are
$\varepsilon _r  = \mu _r  = \frac{{R_2 ^2 }}{{r^2 }}$;
The relative angular electric permittivity and magnetic permeability are
$ \varepsilon _\theta   = \varepsilon _\phi   = \mu _\theta   = \mu _\phi   = \frac{{R_2  - R_1 }}{{r - R_1 }} $.
We discovered new GLHUA analytical method and an exact and analytic radial EM wave solution of the Maxwell equation in the above new GLHUA outer layer annular invisible cloak, $R_1  \le r \le R_2 $,
\[
\begin{array}{l}
 \left[ {\begin{array}{*{20}c}
   {E_r (\vec r)}  \\
   {H_r (\vec r)}  \\
\end{array}} \right] =  \\ 
  = \sum\limits_{l = 1}^\infty  {\left[ {\begin{array}{*{20}c}
   {E_{r,l,1} (\vec r)}  \\
   {H_{r,l,1} (\vec r)}  \\
\end{array}} \right]} \cos \left( {k(R_2  - R_1 )\log \left( {\frac{{r - R_1 }}{{R_2  - R_1 }}} \right)} \right) \\ 
 \sum\limits_{m =  - l}^l {} \left[ {\begin{array}{*{20}c}
   {D_e (\theta ,\phi )}  \\
   {D_h (\theta ,\phi )}  \\
\end{array}} \right]Y_l^m (\theta ,\phi )Y_l^{m*} (\theta _s ,\phi _s ) \\ 
  + \sum\limits_{l = 1}^\infty  {\left[ {\begin{array}{*{20}c}
   {E_{r,l,2} (\vec r)}  \\
   {H_{r,l,2} (\vec r)}  \\
\end{array}} \right]} \sin \left( {k(R_2  - R_1 )\log \left( {\frac{{r - R_1 }}{{R_2  - R_1 }}} \right)} \right) \\ 
 \sum\limits_{m =  - l}^l {} \left[ {\begin{array}{*{20}c}
   {D_e (\theta ,\phi )}  \\
   {D_h (\theta ,\phi )}  \\
\end{array}} \right]Y_l^m (\theta ,\phi )Y_l^{m*} (\theta _s ,\phi _s ). \\ 
 \end{array}
\\\\\\\\\\\\\\\\\\\\\\\\\\\\(GLHUA \\ 1)
\]
In the free space outside of the cloak, $ r > R_2 $,$ E_r (\vec r) = E_{b,r} (\vec r) $ $H_r (\vec r) = H_{b,r} (\vec r)$, on the boundary $ r=R_2 $, $ E_r (\vec r)|_{r = R_2 }  = E_{b,r} (\vec r)|_{r = R_2 } $,$ H_r (\vec r)|_{r = R_2 }  = H_{b,r} (\vec r)|_{r = R_2 } $, and the derivative
${\raise0.7ex\hbox{${\partial E_r (\vec r)}$} \!\mathord{\left/
 {\vphantom {{\partial E_r (\vec r)} {\partial r}}}\right.\kern-\nulldelimiterspace}
\!\lower0.7ex\hbox{${\partial r}$}}|_{r = R_2 }  = {\raise0.7ex\hbox{${\partial E_{b,r} (\vec r)}$} \!\mathord{\left/
 {\vphantom {{\partial E_{b,r} (\vec r)} {\partial r}}}\right.\kern-\nulldelimiterspace}
\!\lower0.7ex\hbox{${\partial r}$}}|_{r = R_2 } $,
${\raise0.7ex\hbox{${\partial H_r (\vec r)}$} \!\mathord{\left/
 {\vphantom {{\partial H_r (\vec r)} {\partial r}}}\right.\kern-\nulldelimiterspace}
\!\lower0.7ex\hbox{${\partial r}$}}|_{r = R_2 }  = {\raise0.7ex\hbox{${\partial H_{b,r} (\vec r)}$} \!\mathord{\left/
 {\vphantom {{\partial H_{b,r} (\vec r)} {\partial r}}}\right.\kern-\nulldelimiterspace}
\!\lower0.7ex\hbox{${\partial r}$}}|_{r = R_2 } $. 
From the above
radial analytical EM wave field, we obtain a new angular analytical EM wave field. The new GLHUA analytical EM
wave field and and incident wave field 
and their radial derivatives are continuous on the outer boundary $r = R_2$. It makes that there is no any scattering from GLHUA outer layer cloak to disturb the incident EM wave field in free space,$r > R_2$. . The above exact and analytical EM wave field can not propagate into, $r < R_1 $, inside of sphere with radius $R_1$, GL no scattering modeling and inversion method can be used to find any
exact and analytical EM full wave solution of the 3D Maxwell equation in
our GLHUA inner layer invisible cloak.
Our GLHUA analytical modeling method can be used to find any
exact and analytical EM full wave solution of the 3D Maxwell equation in
spherical annular layer domain with every invisible cloak materials.
The above (GLHUA 1)  exact and analytical EM wave field  propagation in the new GLHUA outer layer invisible cloak is undisputed evidence to prove that GLHUA and GLLH double layer cloak is a practicable invisible cloak with concealment and with relative EM parameters not less than 1 or with relative refractive index not less than 1 and without exceeding light speed propagation, and the reciprocal principle is satisfied in GLHUA double invisible cloak. The above (GLHUA 1)  exact and analytical EM wave field  propagation in the new GLHUA outer layer invisible cloak is undisputed evidence to prove that GL no scattering modeling and inversion are breakthrough method for discovering invisible cloak and for simulation of full wave field propagation in it . The above (GLHUA 1)  exact and analytical EM wave field  propagation in the new GLHUA outer layer invisible cloak is undisputed evidence to prove that super sciences is really new sciences. The above (GLHUA 1)  exact and analytical EM wave field  propagation in the new GLHUA outer layer invisible cloak is undisputed evidence to prove that the EM wave front and wave ray are discontinuously propagation in GLHUA and GLLH double layer invisible cloak,
The above (GLHUA 1)  exact and analytical EM wave field  propagation in the new GLHUA outer layer invisible cloak [33] is undisputed evidence to expose the software COMSOL's mistakes. 
There are two methods to create Electromagnetic (EM) invisible cloak. One method is our Global and Local EM field no scattering  modeling and inversion. Another method is that 0 to $R_1$ sphere radial transformation method.  In 2000, we discovered double layered EM invisible cloak phenomena [3] in the GILD [2 ] EM modeling and inversion simulations. In 2002, we proposal Global and Local field GL scattering modeling and inversion [5][6] which is powerful method to create EM invisible cloak. In 2006, Pendry et al. used  the linear $0$ to $R_1$ sphere radial transformation to propose the impracticable EM invisible cloak with infinite speed and exceeding light speed propagation physical violation [4]. 
The infinite speed and exceeding light speed propagation are fundamental physical
violation and difficulties that making Pendry type cloaks are impracticable. The fundamental difficulties in Pendry cloak can not be solved by 0 to $R_1$  sphere radial transformation method. After 18 years hard works since 2000, we proposed Global
and Local field GL no scattering modeling and inversion in 2006 and created GLHUA
practicable EM invisible cloaks and their theory [8-9][11-13][15-23][27][29][33][35-39]. 

Recently, we discovered a new negative space first in the world {33} [39], The space in which the radial variable in the sphere coordinate is positive or zero is called positive space, that is real three dimensional space we are living.  The space in which the radial variable in the sphere coordinate is  negative is called negative space. The positive space is visible, but  the negative space is invisible.  The door between the positive and negative is the zero $0$. In the normal sciences, the door $0$ is closed. We proposed the radial 
transform GLHUANPI in this paper and GLHUANPII transformation in my paper 
arXiv:1706.10147 [33] to open the door and translate the negative space to the positive space.  In my paper arXiv:1706.10147 [33] , we discovered a novel sphere radial transformation GLHUANPII that maps the negative infinity, $ - \infty  $  in Basic Space (BS) to $R_1$ in the Physical space (PS) and keep $R_2$ invariant 
\[
r=R_1+(R_2-R_1)e^{(R-R_1)/(R_2-R_1)-1},     ((82) in arXiv:1706.10147[33] )
\]
and its inverse transformation
\[
R=(R_2-R_1)log((r-R_1)/(R_2-R_1))+R_2,     ((83) in arXiv:1706.10147[33] )
\],

that will create new GLHUANPII invisible cloak and GLHUANPII relative EM parameters
\[ 
\varepsilon _\theta   = \mu _\theta   = {{(R_2  - R_1 )} \mathord{\left/
 {\vphantom {{(R_2  - R_1 )} {(r - R_1 )}}} \right.
 \kern-\nulldelimiterspace} {(r - R_1 )}}       ((84) in arXiv:1706.10147[33] )
\],
 
\[
\varepsilon _r  = \mu _r  = \frac{{\left( {(R_2  - R_1 )\log \left( {\frac{{r - R_1 }}{{R_2  - R_1 }}} \right) + R_2 } \right)^2 }}{{r^2 }}\frac{{r - R_1 }}{{R_2  - R_1 }},  ((85) in arXiv:1706.10147[33] )
\]
the electric dielectric and magnetic permeability in angular direction are same as GLHUA EM invisible cloak in my paper arXiv:1706.10147[33]. It is very suprise that
the GLHUANPII transformation can be used to create three different invisible cloaks that are detailed proposed in my paper arXiv:1706.10147[33]  By GL no scattering modeling and inversion and
\[
\varepsilon _\theta   = \mu _\theta   = {{(R_2  - R_1 )} \mathord{\left/
 {\vphantom {{(R_2  - R_1 )} {(r - R_1 )}}} \right.
 \kern-\nulldelimiterspace} {(r - R_1 )}}
\] ,
 we discover the electric dielectric and magnetic permeability in radial direction [33]
	\[
\varepsilon _r  = \mu _r  = {{R_2^2 } \mathord{\left/
 {\vphantom {{R_2^2 } {r^2 }}} \right.
 \kern-\nulldelimiterspace} {r^2 }.}
\]
We obtained the practicable GLHUA outer layered EM invisible cloak without exceeding light speed propagation. That is additional new theoretical base for practicable GLHUA outer layered EM invisible cloak.

The GLHUA analytical modeling and inversion, GLHUA and GLLH double layer invisible cloak and GLHUA exact analytical full EM wave in GLHUA double layer cloak are patent by authors in GL Geophysical laboratory in USA and in Dayuling Super Computation center in China. If anyone colleague use our GLHUA and GLLH double layer invisible cloak and GLHUA exact analytical EM wave propagation for research and product, please be sure to cite our papers  in his papers or reports. otherwise it is essentially to plagiarism  our research results. We hope our practicable GLHUA double layer and GLLH Double layer invisible cloak will be used to invisible cloaking spacecraft for space and interstellar sailing of people in the world We hope our practicable GLHUA double layer and GLLH Double layer cloak will be used for peace and sciences of people in the world. A detailed GLHUA analytical method for EM full wave propagation in GLHUA cloak and GLLH cloak will be published in our next new paper. Copyright and patent of the GLLH EM cloaks and GL modeling and inversion methods, GLHUA analytical modeling and inversion methods, our GLHUA exact analytical EM full wave solution of 3D Maxwell EM equation in GLHUA double cloak and in any sphere annular cloak are reserved by authors in GL Geophysical Laboratory. 
\end{abstract}

\maketitle

\section{\label{sec:level1}INTRODUCTION} 
Since our high school or undergraduate university, we have already been known that
the radial variable in the sphere coordinate is always positive or zero. Recent, in investigation of the wave propagation, first of the world,in my paper [33] and in this paper, we discover the radial variable in the sphere coordinate can be negative. The space in which the radial variable in the sphere coordinate is positive or zero is called positive space, that is real three dimensional space we are living.  The space in which the radial variable in the sphere coordinate is  negative is called negative space. The positive space is visible, but  the negative space is invisible.  In the acoustic equation in the sphere coordinate [32], radial electric wave equation and radial magnetic wave equation in the spherical coordinate [1] [31], if the radial variable $R$ is  changed to $-R$ , the equations are invariant, if the acoustic wave field solution is $u(R,\theta ,\phi )$ , the $u(-R,\theta ,\phi )$  is also solution of the homogeneous acoustic equation. Similar,  if radial electric field $E_r (R,\theta ,\phi )$  and radial magnetic wave $H_r (R,\theta ,\phi )$  are solution of the radial electric and radial magnetic equations, then 
 $E_r (-R,\theta ,\phi )$
and $H_r (-R,\theta ,\phi )$
  are also the solution of the homogeneous radial electric and radial magnetic equations  respectively, Therefore the acoustic equation and radial electric and magnetic equations and their solutions  in the sphere coordinate in the positive space with $R \ge 0$  can be analytical continuation into the negative space with $R < 0$ . That will be proved in the section 2 to section 4. We propose a GLHUANPI transformations in this paper, under  GLHUANPI transformations, the part of the negative space is translated into the other part of the positive space. The wave propagation in that part of negative space is shown in the part of the positive space. 
We propose GLHUANPI transformation and create new GLHUANPI electromagnetic permeability and dielectric in physical whole sphere and find electromagnetic and acoustic wave propagation in the whole sphere,  A new negative space in this paper is an important discoveries in the super sciences and super computational sciences in the world. We combine the negative space and positive space together to be new extended super universe 
$ - \infty  < R <  + \infty $ in sphere coordinate. The door between the positive space and negative space is $0$ .  In normal general sciences excluding religion and theology, the positive space and negative space are not connected. The door 0 between the positive space and negative space is closed. Up to now, all general sciences principles, for example, conservation of energy, are hold and satisfied in the positive space. Up to now, all scientist are working in the positive space.  We create the GLHUANPI and GLHUANPII transformation in [33] to open the door $0$ between the negative and positive space. Nothing is in  
$R <  - \infty$   or  in  $+ \infty  < R $  .The new negative space will have important theoretical and practicable applications in the sciences and super sciences. In the Pendry Cloak. the relative EM parameters are from Pendry transformation in the annular layer $R_1 < R < R_2$ , Also. Pendry put free space media with relative EM parameter $1$ in the inner sphere $0 < R < R_1$ that create Pendry invisible cloak in his paper in sciences 2006[4].In this paper, based on our novel negative space, we discovered and proved an important problem that If the Pendry relative EM  parameters  
are sit in the whole sphere  which include the a nnular  and inner sphere  , then there is nonzero EM wave in the inner sphere  , the inner sphere is not cloaked concealment.

It is totally different from the transformation optics, and completely do not use transformation optics, by using GL no scattering modeling and inversion [1][31], in my paper [33], we discover a new GLHUA outer annular layer EM invisible cloak with relative parameter not less than 1. In GLHUA outer layer invisible cloak , $R_1  \le r \le R_2 $, the relative radial electric permittivity and magnetic permeability are
$\varepsilon _r  = \mu _r  = \frac{{R_2 ^2 }}{{r^2 }}$;
The relative angular electric permittivity and magnetic permeability are
$ \varepsilon _\theta   = \varepsilon _\phi   = \mu _\theta   = \mu _\phi   = \frac{{R_2  - R_1 }}{{r - R_1 }} $.
In [33], we discovered new GLHUA analytical method and an exact analytic radial EM wave solution of the Maxwell equation in the above new GLHUA outer layer annular cloak, $R_1  \le r \le R_2 $, we brief propose here, the detailed derive in [33],
\[
\begin{array}{l}
 \left[ {\begin{array}{*{20}c}
   {E_r (\vec r)}  \\
   {H_r (\vec r)}  \\
\end{array}} \right] =  \\ 
  = \sum\limits_{l = 1}^\infty  {\left[ {\begin{array}{*{20}c}
   {E_{r,l,1} (\vec r)}  \\
   {H_{r,l,1} (\vec r)}  \\
\end{array}} \right]} \cos \left( {k(R_2  - R_1 )\log \left( {\frac{{r - R_1 }}{{R_2  - R_1 }}} \right)} \right) \\ 
 \sum\limits_{m =  - l}^l {} \left[ {\begin{array}{*{20}c}
   {D_e (\theta ,\phi )}  \\
   {D_h (\theta ,\phi )}  \\
\end{array}} \right]Y_l^m (\theta ,\phi )Y_l^{m*} (\theta _s ,\phi _s ) \\ 
  + \sum\limits_{l = 1}^\infty  {\left[ {\begin{array}{*{20}c}
   {E_{r,l,2} (\vec r)}  \\
   {H_{r,l,2} (\vec r)}  \\
\end{array}} \right]} \sin \left( {k(R_2  - R_1 )\log \left( {\frac{{r - R_1 }}{{R_2  - R_1 }}} \right)} \right) \\ 
 \sum\limits_{m =  - l}^l {} \left[ {\begin{array}{*{20}c}
   {D_e (\theta ,\phi )}  \\
   {D_h (\theta ,\phi )}  \\
\end{array}} \right]Y_l^m (\theta ,\phi )Y_l^{m*} (\theta _s ,\phi _s ). \\ 
 \end{array}\
\]
In the free space outside of the cloak, $r > R_2 $,$E_r (\vec r) = E_{b,r} (\vec r)$ $H_r (\vec r) = H_{b,r} (\vec r)$, on the boundary $r=R_2$, $E_r (\vec r)|_{r = R_2 }  = E_{b,r} (\vec r)|_{r = R_2 } $,$H_r (\vec r)|_{r = R_2 }  = H_{b,r} (\vec r)|_{r = R_2 } $, and the derivative
${\raise0.7ex\hbox{${\partial E_r (\vec r)}$} \!\mathord{\left/
 {\vphantom {{\partial E_r (\vec r)} {\partial r}}}\right.\kern-\nulldelimiterspace}
\!\lower0.7ex\hbox{${\partial r}$}}|_{r = R_2 }  = {\raise0.7ex\hbox{${\partial E_{b,r} (\vec r)}$} \!\mathord{\left/
 {\vphantom {{\partial E_{b,r} (\vec r)} {\partial r}}}\right.\kern-\nulldelimiterspace}
\!\lower0.7ex\hbox{${\partial r}$}}|_{r = R_2 } $,
${\raise0.7ex\hbox{${\partial H_r (\vec r)}$} \!\mathord{\left/
 {\vphantom {{\partial H_r (\vec r)} {\partial r}}}\right.\kern-\nulldelimiterspace}
\!\lower0.7ex\hbox{${\partial r}$}}|_{r = R_2 }  = {\raise0.7ex\hbox{${\partial H_{b,r} (\vec r)}$} \!\mathord{\left/
 {\vphantom {{\partial H_{b,r} (\vec r)} {\partial r}}}\right.\kern-\nulldelimiterspace}
\!\lower0.7ex\hbox{${\partial r}$}}|_{r = R_2 } $. 
Detailed proof is proposed in our new paper [33].
From the above
radial analytical EM wave field, we obtain a new angular analytical EM wave field. The new GLHUA analytical EM
wave field and and incident wave field 
and their radial derivatives are continuous on the outer boundary $r = R_2$. It makes that there is no any scattering from GLHUA outer layer cloak to disturb the incident EM wave field in free space,$r > R_2$. . The above exact and analytical EM wave field can not propagate into, $r < R_1 $, inside of sphere with radius $R_1$, GL no scattering modeling and inversion method can be used to find any
exact and analytical EM full wave solution of the 3D Maxwell equation in
our GLHUA inner layer invisible cloak.
Our GLHUA analytical modeling method can be used to find any
exact and analytical EM full wave solution of the 3D Maxwell equation in
spherical annular layer domain with every invisible cloak materials.

Our GLHUA analytical modeling method can be used to find any
exact analytical EM full wave solution of the 3D Maxwell EM equation in
our GLHUA inner cloak.
Our GLHUA analytical modeling method can be used to find any
exact analytical EM full wave solution of the 3D Maxwell EM equation in
spherical annular layer domain with every invisible cloak materials.

The above (GLHUA 1)  exact and analytical EM wave field  propagation in the new GLHUA outer layer invisible cloak is undisputed evidence to prove that GLHUA and GLLH double layer cloak is a practicable invisible cloak with concealment and with relative EM parameters not less than 1 or with relative refractive index not less than 1 and without exceeding light speed propagation, and the reciprocal principle is satisfied in GLHUA double invisible cloak. The above (GLHUA 1)  exact and analytical EM wave field  propagation in the new GLHUA outer layer invisible cloak is undisputed evidence to prove that GL no scattering modeling and inversion are breakthrough method for discovering invisible cloak and for simulation of full wave field propagation in it . The above (GLHUA 1)  exact and analytical EM wave field  propagation in the new GLHUA outer layer invisible cloak is undisputed evidence to prove that super sciences is really new sciences. The above (GLHUA 1)  exact and analytical EM wave field  propagation in the new GLHUA outer layer invisible cloak is undisputed evidence to prove that the EM wave front and wave ray are discontinuously propagation in GLHUA and GLLH double layer invisible cloak,
The above (GLHUA 1)  exact and analytical EM wave field  propagation in the new GLHUA outer layer invisible cloak is undisputed evidence to expose the software COMSOL's mistakes. The GLHUA analytical modeling and inversion, GLHUA and GLLH double layer invisible cloak and GLHUA exact analytical full EM wave in GLHUA double layer cloak are patent by authors in GL Geophysical laboratory in USA and in Dayuling Super Computation center in China. If anyone colleague use our GLHUA and GLLH double layer invisible cloak and GLHUA exact analytical EM wave propagation for research and product, please be sure to cite our papers  in his papers or reports. otherwise it is essentially to plagiarism  our research results. We hope our practicable GLHUA double layer and GLLH Double layer invisible cloak will be used to invisible cloaking spacecraft for space and interstellar sailing of people in the world We hope our practicable GLHUA double layer and GLLH Double layer cloak will be used for peace and sciences of people in the world. A detailed GLHUA analytical method for EM full wave propagation in GLHUA cloak and GLLH cloak will be published in our next new paper. A extremely novel  and surprising GLHUA mirage [13] will be published. Copyright and patent of the GLLH EM cloaks and GL modeling and
inversion methods, GLHUA analytical modeling and inversion methods, our GLHUA exact analytical EM full wave solution of 3D Maxwell EM equation in GLHUA double cloak and in any sphere annular cloak are reserved by authors in GL Geophysical Laboratory.
Recently, some colleague comment: why did you define radial GL EM field in (18) in arXiv:1701.02583? We answer that because in 3D spherical coordinate system, the radial electromagnetic equation  (5) and (6) in arXiv1701.02583 are different from the acoustic equation. The equation (5) and (6) are not self-conjugate operator equation. After definition of our GL radial EM field in equation (18) in arXiv:1701.02583, our GL radial EM equation (19) and (23) in arXiv:1701.02583 are self-conjugate equation. Moreover, in $0$ to $R_1$ radial coordinate transformation, the radial EM wave field are changed. Our GL radial electric wave field $E(\vec r) = r^2 \varepsilon _r (r)E_r (\vec r)$, and GL radial magnetic wave field $H(\vec r) = r^2 \varepsilon _r (r)H_r (\vec r),$ are invariant. In free space, when $r$ is going to zero, the GL radial EM field are going to zero. Based on diefination of the GL radial EM field, in the physical annular layer, $R_1 \le r_q \le R_2$, when $r_q$ is going to $R_1$, the magnetic flux and electric displacement are going to zero. Therefore, the $0$ to $R_1$ radial coordinate transformation can be used for making EM invisible cloak with infinite speed and exceeding light speed phase velocity fundamental difficulties. However, in spherical coordinate and in the $0$ to $R_1$ radial coordinate transformation, $r_q  = R_1  + Q(r)$   , suppose that there is no scattering acoustic wave from the sphere $r \le R_2 $ to disturb the incident pressure acoustic wave field, $p_i(r)$, in outside sphere , $r > R_2 $ , then on the inner spherical boundary of physical cloak, the pressure acoustic wave field $p(R_1 ) =  - \frac{1}{{4\pi }}\frac{{e^{ik_b r_s } }}{{r_s }},$. Again suppose that there is no scattering acoustic wave from the sphere $r \le R_1 $ to disturb the pressure acoustic wave field 
$p(r),R_1  \le r \le R_2 $, then there is nonzero spherical symmetry pressure acoustic wave propagation in the inner sphere $r \le R_1 $,
\[
\begin{array}{l}
 p(r) =  - \frac{{R_1 ^2 k_b^2 }}{{4\pi }}\frac{{e^{ik_b r_s } }}{{r_s }} \\ 
 \left( {n'_0 (k_b R_1 )j_0 (k_b r) - j'_0 (k_b R_1 )n_0 (k_b r)} \right), \\ 
 \end{array} \\ (GLHUA \ 2),
\]
sphere $r< R_1$ can not be cloaked. Therefore the $0$ to $R_1$ radial coordinate transformation can not be used for making acoustic cloak. The detailed proof is published in my new arXiv paper ��Problem on the acoustic cloak by 0 to R1 transformation�� with arXiv:submit/1920322 in June 14, 2017. First in the world , we proposed and make clear that 0 to R1 radial transformation can not be used for making acoustic cloak.. 

In this paper, we discovered a class of GLHUA electromagnetic (EM) invisible double layer cloak with relative EM parameter not less than 1 for each layer, and each layer with any thickness. Our idea and method to create GLHUA cloak is different from other cloak. Using our GILD and GL no scattering modeling and inversion, we find nonzero solution of no scattering nonlinear EM inversion equation to create a novel GLHUA EM invisible double layer cloak. We proved that GLHUA cloak is complete invisible cloak with concealment, and phase velocity of EM wave is less than light speed and tends to zero at boundary  in the cloak. The rigorous and detailed theoretical proof is proposed in our paper [1] to support this paper. In 2000, in GILD EM modeling and inversion simulation [2], we observed a strange double layer cloak phenomena in Figure 1 (Fig.2 in [3]). The double layer cloak phenomena was called GILD phenomena that has been published in SEG Expanded Abstracts, Vol. 21, No. 1, 692-695, 2002, by Jianhua Li at al. [3]. The double layer GILD phenomena cloaking imaging in Figure 1 and 2 is a nonzero solution image of no scattering inversion which cloaked the solution image of scattering inversion. Look like general solution of linear operator equation which is the sum of nonzero solution of the homogeneous equation plus the special solution of the inhomogeneous equation. The nonzero solutions of the homogeneous linear equation form a linear space. From non uniqueness of scattering inversion, the nonzero solutions of the no scattering nonlinear inversion equation form more complicated class structure. We developed Global Integral and Local Differential GILD and Global and Local GL field no scattering modeling and inversion to create GLHUA double cloak. After 16 year hard research works, in this paper, GLHUA invisible EM double layer cloak is a novel model and theory for explaining the double layer cloak phenomena discovered in 2000 [3]. By linear coordinate transform in optics, Pendry at al proposed invisible cloak in Science in 2006 [4] with relative radial parameter less than 1 and tends to zero that making the phase velocity of the wave in Pendry cloak is exceeding light speed and tends to infinity and reciprocal is not satisfied. It is different from Pendry cloak [4], GLHUA double layer cloak is created by GL no scattering modeling and inversion [5-6]. The phase velocity of EM wave propagation in GLHUA cloak is less than light speed and tends to zero; but the phase velocity in Pendry cloak is exceeding light speed and tends to infinite; The reciprocal principle is satisfied in GLHUA double layer cloak; but the reciprocal principle is not satisfied in Pendry cloak; The EM material of GLHUA double layer cloak with relative parameter not less than 1 that can be find in natural world, and GLHUA double layer cloak is practicable; but the radial relative parameter and relative refractive parameter are less than 1 in Pendry cloak, the material of Pendry cloak can not be find in natural world and can not be practicable. GLHUA double cloak does overcome the above three fundamental difficulties in Pendry cloak and has more advantages than Pendry cloak. From Pendry cloak paper published in Science in 2006, many research papers to follow Pendry cloak and optical transform method were published. However, up to now, the above fundamental difficulties in Pendry cloak have no been solved, and the three difficulties can not be solved by the optical transform. GLHUA double layer cloak and GL no scattering inversion method [5-6] overcome the three fundamental difficulties in Pendry cloak that is breakthrough and open a new research door for invisible cloak. In paper by Pendry et al in Science in 2006 [4], Pendry stated that nor can any radiation get out. that is totally wrong. We proved that "any radiation excited in Pendry cloak concealment will be propagation go to out." Zhang and Chen at al in [7] proved that there exist extraordinary surface Voltage effects in Pendry cloak with an active device inside, that is as same as that we proved in [8][16]. In [8], to that nor can any radiation get out as Pendry stated in [4], we proved a contradiction conclusion that no Maxwell EM wave field can be excited in cloaked concealment. The contradiction inversely does prove that any radiation excited in Pendry cloaked concealment can propagation to go to outside of the cloak. Reciprocal principle is not satisfied in Pendry cloak. For overcoming this difficulty, we proposed novel GLHUA inner cloak in this paper and double layer cloak in paper [9]. Up to now, only our inner layer cloak in double layer cloak are proposed [9] [17-22] that making the reciprocal principle is satisfied and the EM environment in concealment is not be disturbed by cloak. The ULF cloak [10] does overcome infinite speed difficulty, but in ULF cloak [10], the phase velocity of the wave is exceeding light speed and reciprocal principle is not satisfied. In May of 2010, G. Xie et al. published GLLH EM Invisible Cloak with Novel Front Branching and without Exceed Light Speed Violation in arXiv1005.3999 [11]. The paper also was presented in PIERS 2010 [12]. The three fundamental difficulties in Pendry cloak and the two difficulties in ULF cloak are overcome in GLLH cloak [11] and double layer cloak [9] [17-22]. The phase velocity of GLLH cloak is less than light speed and tends to zero in the inner boundary of the outer layer cloak. A New computational mirage has been published in PIERS abstract in Hang Zhou of China by F. Xie and Lee Xie, 296 in 2005 [13]. Next year, in 2011, Ulf Leonhardt et al. published paper Invisibility cloaking without superluminal propagation in arXiv: 1105.0164v3 [14]. Ulf did write a review and references of cloak history in his paper [14], he cited many papers about Pendry cloak and all published cloak papers and wrote that The fundamental problem is that perfect invisibility (Pendry cloak [4] )requires that light should propagate in certain cloaking regions with a superluminal phase velocity that tends to infinity. In paper [14] in 2011, Ulf cited our paper arXiv: 1005.3999 in 2010 as his reference [35] and wrote that The preprint [35], proposes a different method for cloaking without superluminal propagation.In same paper, ULF proposed their Invisibility cloaking without superluminal propagation. Up to now, there is no full wave theoretical analysis and no full wave computational simulation to verify ULFs cloak. From 2010 to now, we again expense 6 year for deep research, in this paper, we discovered the new GLHUA double layer cloak with relative EM parameter Not Less Than 1 for any thickness double annular layers. In the outer annular layer cloak,$R_1  \le r \le R_2 $, the relative electric permittivity
\[
\varepsilon _r  = 1,\varepsilon _\theta   = \varepsilon _\phi   = \frac{1}{2}\left( {\left( {\frac{{r - R_1 }}{{R_2  - R_1 }}} \right)^\alpha   + \left( {\frac{{R_2  - R_1 }}{{r - R_1 }}} \right)^\alpha  } \right)
\]
and the relative magnetic permeability $\mu _r  = \varepsilon _r $, 
$\mu _\theta   = \mu _\phi   = \varepsilon _\theta  $,
In the inner layer cloak $R_0  \le r \le R_1 $, relative permittivity

\[
\varepsilon _r  = 1,\varepsilon _\theta   = \varepsilon _\phi   = \frac{1}{2}\left( {\left( {\frac{{R_1  - r}}{{R_1  - R_0 }}\frac{{R_0 }}{r}} \right)^\alpha   + \left( {\frac{{R_1  - R_0 }}{{R_1  - r}}\frac{r}{{R_0 }}} \right)^\alpha  } \right),
\]
The other place, $r \le {R_0} $  or  $r \ge {R_2 }$ , is free space.
and the relative magnetic permeability
$\mu _r  = \varepsilon _r $, 
$\mu _\theta   = \mu _\phi   = \varepsilon _\theta  $,
$0 < \alpha _0  \le \alpha  \le \alpha _1  < 2$,
In the paper [1], in our GLHUA double layers cloak in any thickness inner layer,
$\left[ {R_0 ,R_1 } \right]$, and outer layer $\left[ {R_1 ,R_2 } \right]$,

for $\alpha  = 1$, and all frequency, by using our Global and Local (GL) EM full wave modeling and no scattering inversion theorem, we created GLHUA double layer cloak and rigorous proved that when r going to boundary, $r = R_1 $, the EM wave field is going to zero; we proved that all incident EM wave excited in outside of cloak can not be disturbed by cloak, and the EM wave can not propagation penetrate into the concealment. Reciprocally, incident EM wave excited in inside of concealment can not be disturbed by cloak, and the EM wave can not propagation to outside of cloak. Our GLHUA cloak is practicable. The detailed and rigorous theoretical proof is presented in paper [1] with 8 sections. In the section 2 in [1], we proved that the relative parameter in (1) (in (1) in [1]) and in (2) (in (12) in [1]) and their derivative are continuous across the outer boundary $r = R_2 $,$r=R_0$ , proved that the relative parameter not less than 1. In section 3 in [29] we define the radial GL EM field, $E(\vec r)$,$H(\vec r)$,in (18) ; Proposed GL EM second order radial field differential equation (19) and (23) in [29], the radial GL EM field are invariance under coordinate transform, 
$E'(\vec r') = E(\vec r)$, $H'(\vec r') = H(\vec r)$:
Discovered and proved new essential property that with weight $\sin \theta $, the spherical surface integral of GL radial wave field equal to zero; Proposed GL adjoint Green field equation and GL Green differential equation; Proposed the radial GL integral equation (36). In section 6 in [29], we proposed GLHUA pre cloak material conditions $(6.1) \ to \ (6.4) $in invisible virtual sphere $r \le R_2 $. In theorem $6.1 \ to \ 6.4$, based on the GHUA pre cloak conditions, we rigorous proved that GL EM wave excited in outside of the cloak smoothly propagation enter to the invisible virtual sphere and going to zero when r going to zero. These theorems are detailed proved by using GL modeling and LHOPITAL ROLE in section 6 the paper [29]. We proposed a new GLHUA angular Green equation (114) in [29] and GLHUA angular Green Function in (115) in [29] that is different from GL Green equation (19), (23) in [29] and GL Green function (18) in [29]. In section 4 in [1], using GL no scattering inversion, we create the GHUA outer layer cloak. Using GL modeling and LHOPITAL Rule, we proved that in outside the cloak the EM wave propagation smoothly enter outer annular layer cloak and going to zero at inner spherical annular $r = R_1 $. In section 5 in [1], using GL no scattering modeling and inversion, we prove that the EM wave field excited in outside of the GHUA cloak propagation can not penetrate to the concealment. In outside the cloak, the EM wave field can not be disturbed by the cloak. In outside the cloak, the total EM field equal to its incident wave field. The whole cloak is invisible. In section 6 in [1], using GL no scattering modeling and inversion, we prove that the EM wave field excited in GHUA cloak concealment can not propagation go to outside of the inner cloak. The EM wave field excited in the concealment can be not disturbed by the cloak. Using GL no scattering modeling and inversion, we create GLHUA double layer cloak and rigorously proved that GLHUA double layer cloak is invisible cloak with concealment. The relative parameter of GLHUA cloak is not less than 1. The EM wave is propagating through GLHUA cloak without infinite speed and without exceeding light speed propagation. Reciprocal principle is satisfied in GLHUA double layer cloak. GL EM Eikonal equation for anisotropic material and for GLHUA cloak is proposed in section 7 in [1]. The idea of creating GLHUA cloak model, GL no scattering modeling and inversion method, full wave theoretical proof analysis and computational simulation by GL method are new and different from all other cloak publications by other researchers. Up to now, there is no other researcher to propose No scattering modeling and inversion. We proposed GL no scattering and inversion in [3] [5] [6] [11] [15].
I, an American Chinese female senior and sick scientist, 50 years to learn Chairman Mao's inversion philosophy idea, suffered and silent  to create and invent, created and developed 3D finite element method and software first in China . and discovered the superconvergent of 3D FEM first in the world and  created GL and GILD scattering and no scattering modeling and inversion etc dozen major novel methods achievements.In particular, from we discovered "GILD phenomena" double layer invisible cloak in 2000, I led my research team after 17 years of extremely hard work and creation  to get today's breakthrough results of creating our GLHUA double layer invisible cloaks with relative parameters not less than 1. We have found  an exact full EM wave propagation solution of the EM equation with the practicable GLHUA double layer perfect invisible (stealth) cloak , which is great breakthrough progress. The exact solution of EM equation with GLHUA and GLLH double layer cloak is indisputable evidence to prove that our GLHUA GLLH double layer invisible cloak model is completely  practicable perfect invisible cloak physics mathematical model. The exact EM wave propagation in the GLLH and GLHUA cloak is published in the version 5 of this paper  .
During 17 years, under I lose job and have no any salary and no any fund, suffered extremely life difficulties and severe pain, caused by over load of GLHUA inversion and GLHUA double invisible cloak and
caused by some one uncreasonable blcoked our paper publication and caused by
unreasonable academic corruption (Recently, Huang Yunqing and Chen Chuanmiao did plagiarism  our 
research result "First in the world, in 1973, we  found three-dimensional finite element super-convergence"[26][34]. , Chen and Huang lie to request Chinese Natural Science Award. Their falsification is based on Chen Chuanmiao's infamous rumors "1977, Chen Chuanmiao et al found multi-dimensional finite element superconvergence in the arch dam calculation." . Because the observation of superconvergence must compare the exact solution and finite element approximation of displacement and stress of the arch dam, and Chen, Chuanmiao and anyone can not know the real displacement and stress of the actual arch dam. All Huang, Yunqing and Chen, Chuanmiao's rumors of the report is very obvious lie. What is moreover, Zhejiang University's Gang Bao does include our scattering forward and inversion research results[24][25][30] as his results lie to request Chinese Natural Science Award. We have already reported the above two academic corruptions to Chinese government). Because Pendry and ULF have no inverse scattering and "No scattering Inversion" knowlegements and concept, they use 
transformation optics to study invisible cloak. In 2009, Pendry reject our paper
because he did not understand our GL modeling and inversion. his review is mistake. I hope Pendry change mind after he read our new papers and GLHUA double layer invisible cloak and exact and analytical EM wave solution of 3D full wave MAXWEll equation in our new GLHUA outer layer invisible cloak. Welcom all Professor and Professor Pendry and ULF to read our paper and give fair comment. However  Gunther Uhlnamm does not know any "zero scattering inversion". Gunther Uhlnamm is not qualified to hold and review our paper.  I am very angry and strongly oppose Gunther Uhlnamm 
to hold and block our paper publication.  If anyone colleague use our GLHUA and GLLH double layer invisible cloak and GLHUA exact and analytical EM wave propagation for research and product, please be sure to cite our papers in his papers and reports.  otherwise it is essentially to plagiarism our research results.
We using our created GILD and GL scattering/no scattering modeling and inversion method to discover and create GLLH and GLHUA double layer invisible  cloak, process theoretical proof and perform  full wave simulation for EM wave propagation in our GLLH and GLHUA cloak.  We , first in the world, proposed the following concepts, methods and terms that: "No scattering modeling and inversion", "no scattering inversion", "no scattering inversion and cloak", and "inversion and cloak". Our GL double layer cloak and GLLH double layer cloak and GLHUA double layer cloak are discovered and created by our GILD and GL no scattering modeling and inversion that is totally different from Pendry cloak and ULF cloak.
The 6 years after we discovered "GILD phenomena" double layer invisible cloak in 2000, in 2006, J.B. Pendry proposed first invisible cloak[4], the Pendry cloak with infinite speed and exceeding light speed was created by Pendry's linear space expansion transformtion optics. ULF cloak was proposed in 2009[10], which with exceeding light speed and without full wave theoretical proof and without full wave simulation verification. ULF cloak was created by ULF's space expansion transformation optics. There is no "inverse Problem" or "inversion" In Pendry and ULF cloak papers.
In our paper arXiv:0904.3168 in April of 2009[9],  we did write "Using the 3D GL EM modeling [1-2] and GL inversion [3], we propose an EM double layered cloak in this paper which is called as GL double layered cloak."
in May of 2010, We published "GLLH EM Invisible Cloak With Novel Front Branching And Without Exceed Light Speed Violation" in  arXiv:1005.3999[11],
In the paper, we did write that "Our GLLH cloak is created by GL EM 
modeling and GL EM cloak inversion with searching class.....".. We named our cloak as GLLH cloak.
There is substantial evidence to show that after Gunther Uhlmann review and unreasonably rejected our paper arXiv:0904.3168 in April of 2009[9] and read our paper arXiv:1005.3999 [11] in May of 2010, 
After Gunther Uhlmann did know that we use GILD and GL No scattering inversion to create our GLLH double layer cloak,
then From 2011, Gunther Uhlmann based on our "inversion and cloak" idea and concepts in our papers, he talks "Inverse Problem and Harry Poter Cloak " in USA, Hong Kong ,and every where in China, but Gunther Uhlmann  did not cite and mention  our cloak and our inversion method  in his talks. In Gunther Uhlmann's papers and talks on "inverse Problem", He did not cite our pioneer papers on inversion of the coefficient of the 3D wave equation [24]]25] in Chinese sciences in 1988 and paper [30] in 1986 in "communication on pure math. and applied math.". 
In February 28, 2017, Gunther Uhlmann gave speak on "Inverse Problems and Harry Potter's Cloak" in central south University in Changsha.see
http://en.csu.edu.cn/info/1084/1770.htm
 It is very interesting that in March 6, 2016, in same university, we gave speak "Demon talking about inversion and GLHUA invisible cloak" in central south University in Changsha. see, http://news.csu.edu.cn/info/1004/126458.htm and figure 18,
In our speak, we proposed "Super Sciences" which has been published in version 1 of this paper. submitted in December 1 of 2016. Based on Chairman Mao's inversion idea, we proposed "SUPER SCIENCES" that "In current general science, the visible (scatterin) natural science and invisible (no scattering) thinking and invisible  social science are main object. Inversely, the invisible (No Scattering) natural science and visible (scttering) thinking and visible social science are main object of the super sciences. Our GLHUA practicable double layer invisible cloak and Easton LaChappelle's mind controlled hands of robot
show that the new super science is being born. Supercomputational sciences are entrance  and channel for studying super sciences. Based on our speak, Hunan Province organized "Super Computational Sciences Society in Hunan Province" in Changsha in 2016.
By Comparison between our speak and  Gunther Uhlmann's speak, all Professor and Student and Hunan People said "Chairma Mao really is great inversion mentor". By Comparison between our speak and  Gunther Uhlmann's speak, Hunan will develop "Chinese Dayuling Supercomputational Sciences Center".By comparison between our speak and Gunther speak, it is proved that practicable GLHUA double layer cloak is breakthrough progress in the invisible super sciences. Gunther
Uhlmann lags behind us for a scientific era.

  $\boldsymbol {CONTENTS}$  \\
I. \ \ Introduction \\
II. \  \ GLHUA EM INVISIBLE DOUBLE LAYERS\\
\  \ CLOAK MATERIALS AND PROPERTIES \\
III.\  \  CREATE GLHUA DOUBLE LAYER CLOAK\\
IV.\  \  GL NO SCATTERING MODELING\\
\  \ METHOD IS USED FOR SIMULATION OF THE\\
\  \ EM WAVE PROPAGATION THROUGH GLHUA\\
\  \ OUTER LAYER CLOAK  \\
V.\  \  COMPARISON BETWEEN GL FULL\\
\  \ WAVE NO SCATTERING MODELING METHOD\\
\  \ AND RAY TRACING METHOD FOR CLOAK \\
VI.\  \  THE RADIAL IN THE SPHERE\\
\  \ COORDINATE CAN BE NEGATIVE AS WELL\\
\  \ AS A NEW NEGATIVE SPACE IN WHICH\\
\  \ ACOUSTIC AND ELECTROMAGNETIC WAVE\\
\  \ PROPAGATION  \\
VII. \  \  DISCUSSION AND CONCLUSION\\

\begin{figure}[b]
\centering
\includegraphics[width=0.86\linewidth,draft=false]{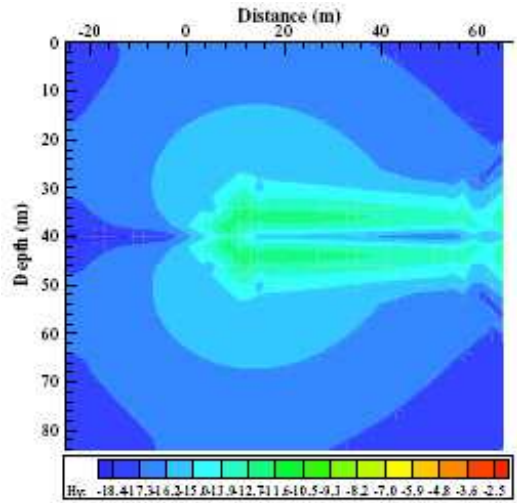}
\caption{ (color online) 
This is figure 2 in reference [3], we observed a strange double layer cloak phenomena in GILD modeling and inversion for shuttle model In 2000, which was published in SEG expand abstract in 2002[]
.}\label{fig1}
\end{figure}
\begin{figure}[b]
\centering
\includegraphics[width=0.86\linewidth,draft=false]{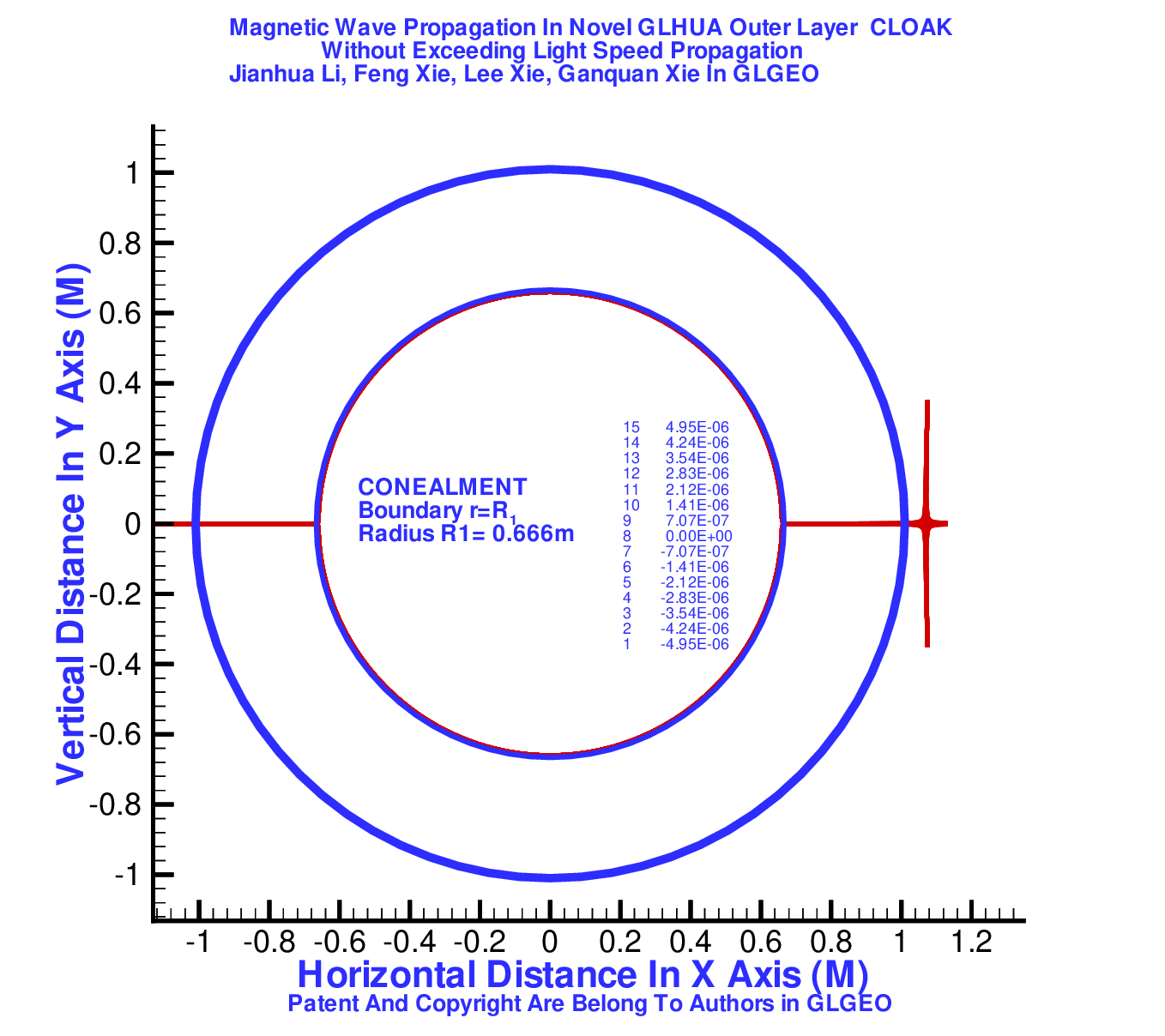}
\caption{ (color online) 
Magnetic wave propagation through GLHUA 
outer layer cloak, wave front at 4 step.}\label{fig5}
\end{figure}
\begin{figure}[b]
\centering
\includegraphics[width=0.86\linewidth,draft=false]{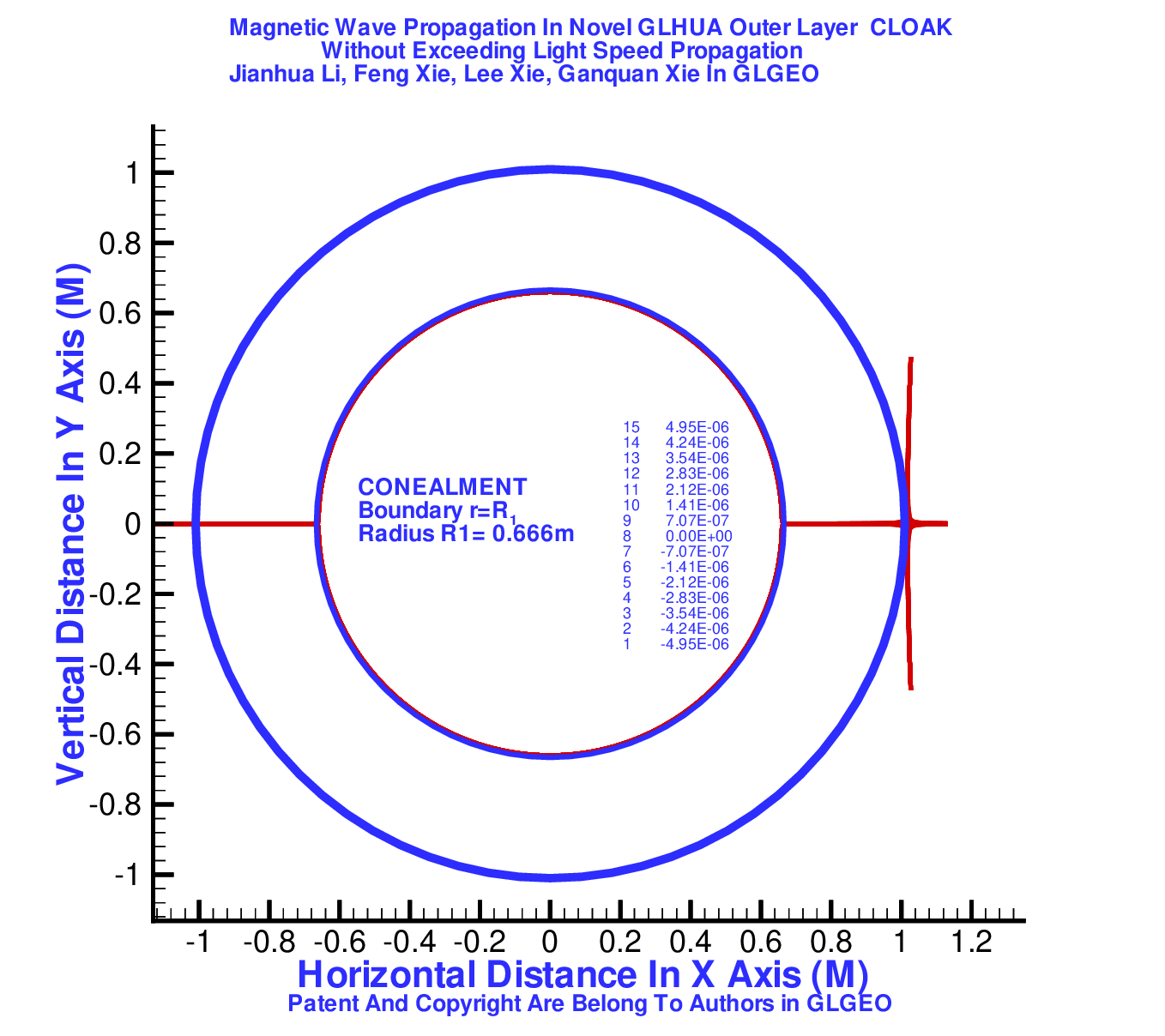}
\caption{ (color online)
Magnetic wave propagation through GLHUA 
outer layer cloak, wave front at 5 step. }\label{fig6}
\end{figure}
\begin{figure}[b]
\centering
\includegraphics[width=0.86\linewidth,draft=false]{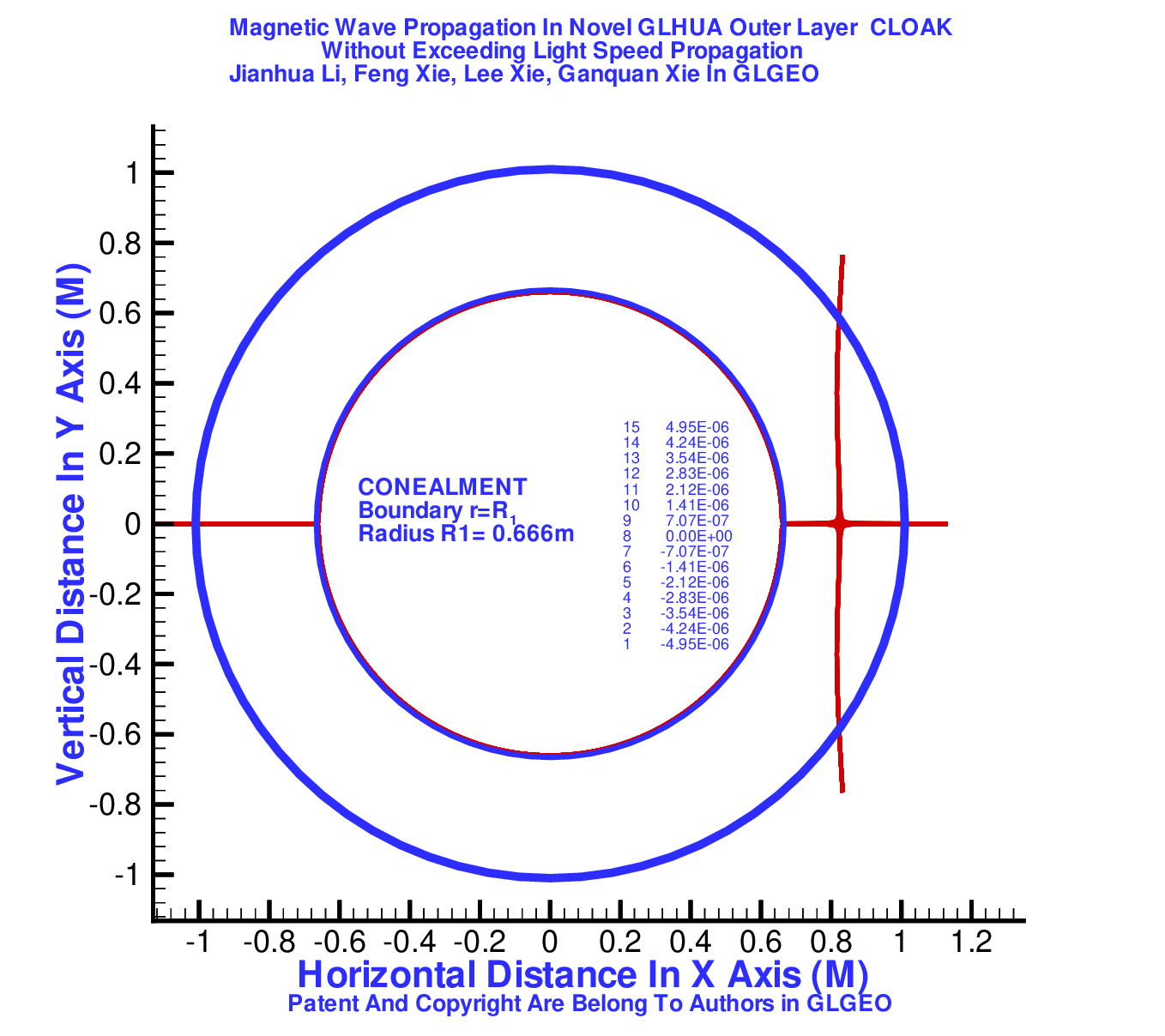}
\caption{ (color online)
Magnetic wave propagation through GLHUA outer layer cloak, wave front at 9 step. 
.}\label{fig7}
\end{figure}
\begin{figure}[h]
\centering
\includegraphics[width=0.86\linewidth,draft=false]{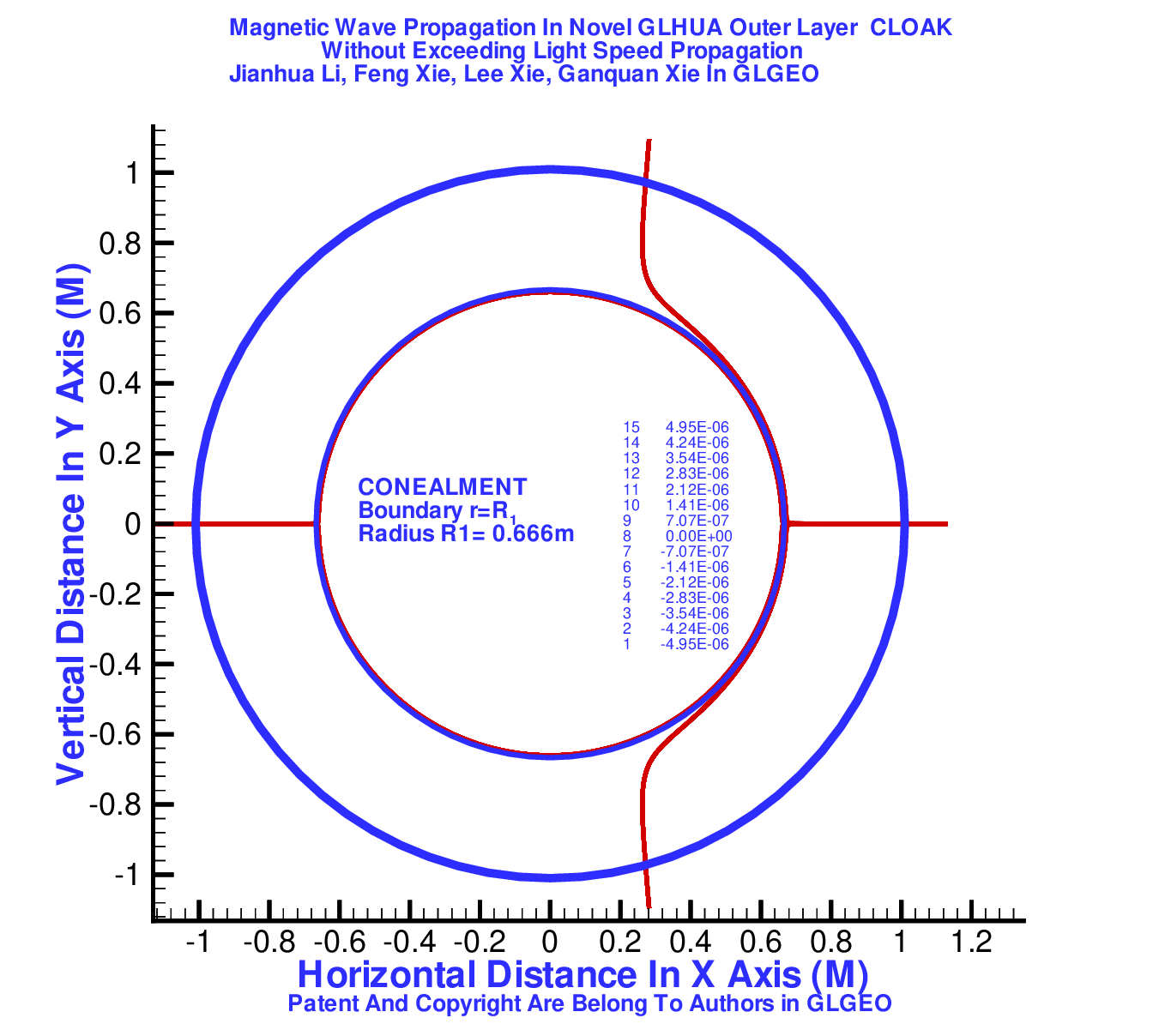}
\caption{ (color online) 
Magnetic wave propagation through GLHUA outer layer cloak, wave front at 20 step.}\label{fig8}
\end{figure}
\begin{figure}[h]
\centering
\includegraphics[width=0.85\linewidth,draft=false]{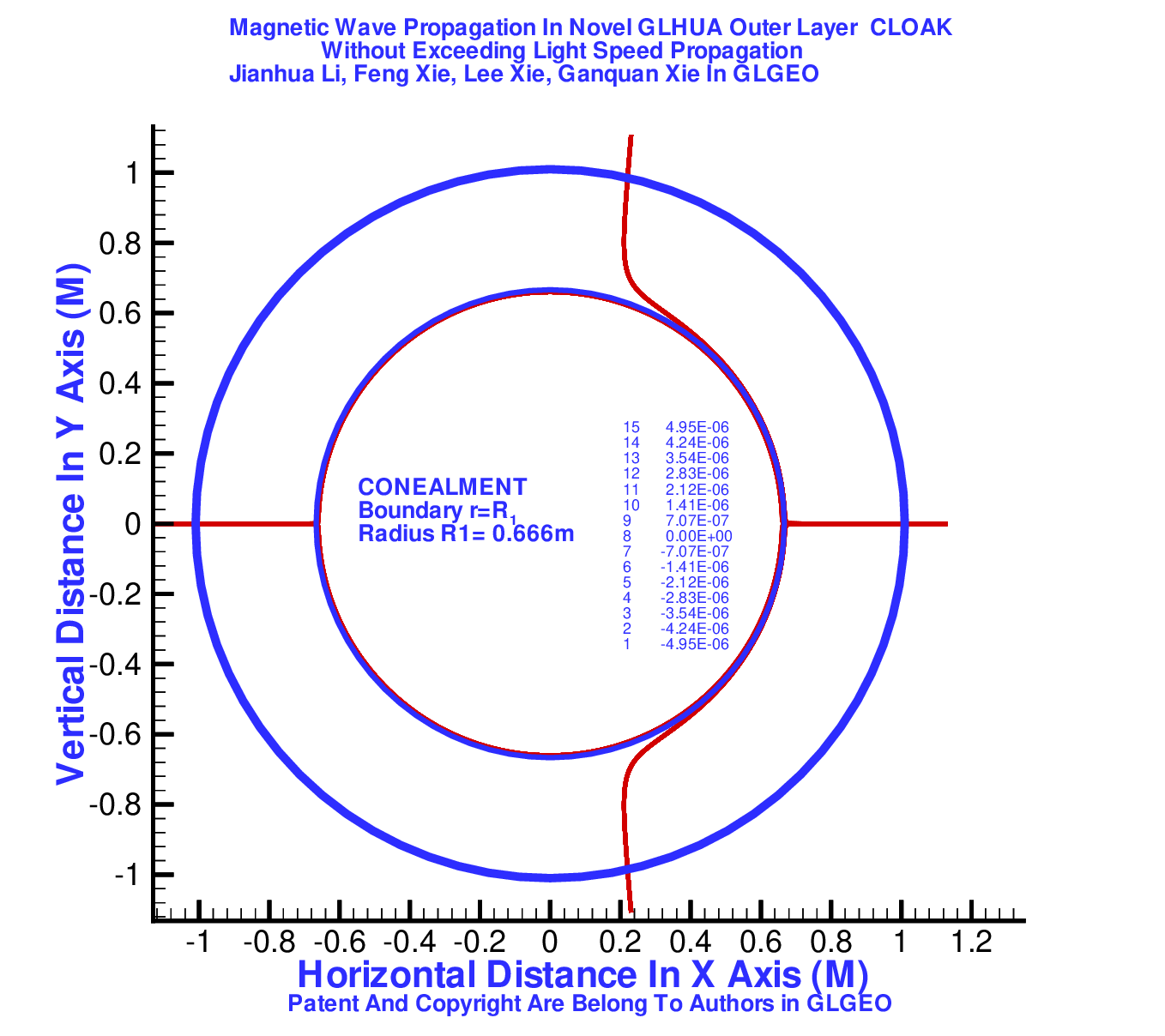}
\caption{ (color online) 
Magnetic wave propagation through GLHUA outer layer cloak, wave front at 21 step.}\label{fig9}
\end{figure}
\begin{figure}[h]
\centering
\includegraphics[width=0.85\linewidth,draft=false]{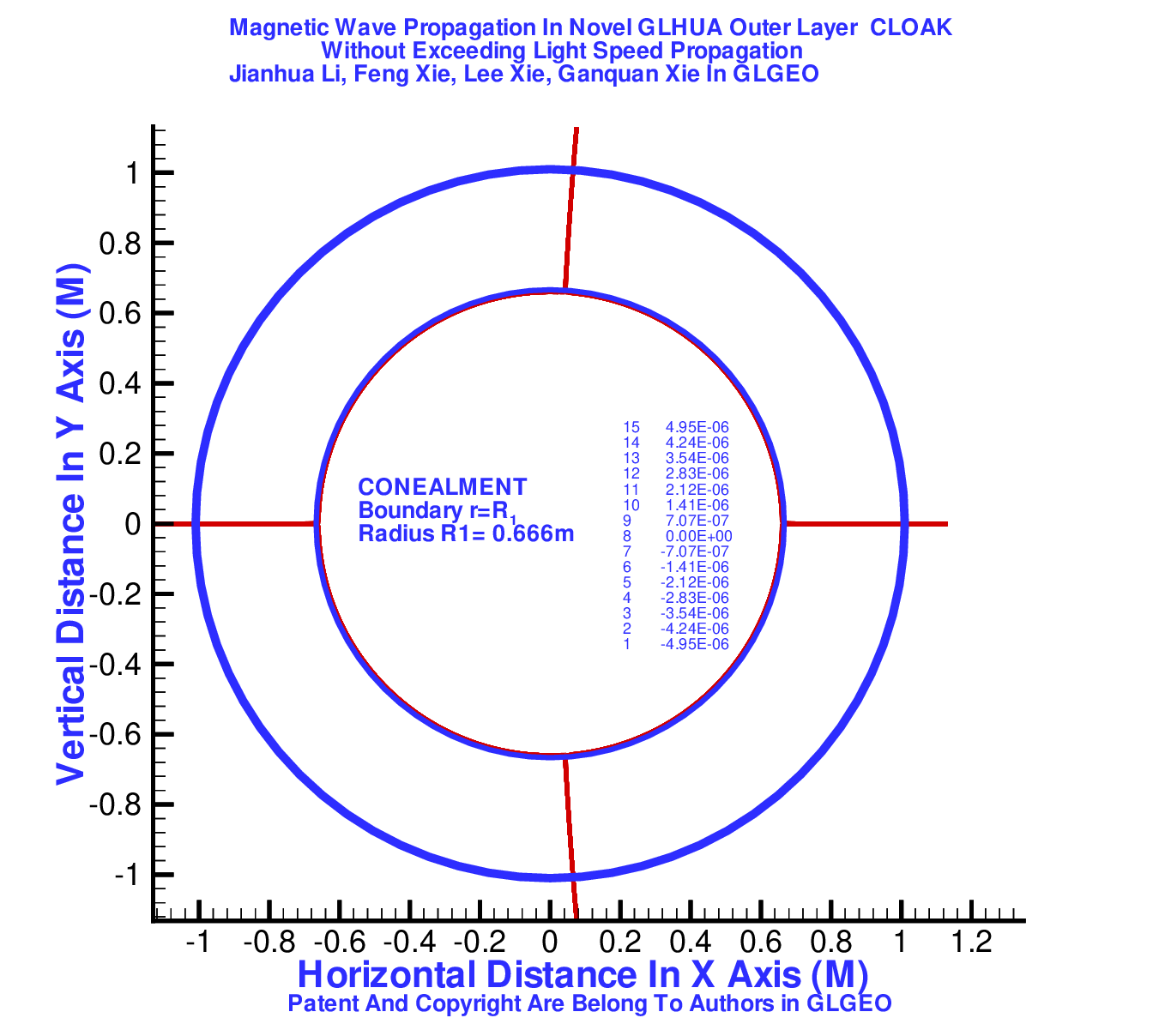}
\caption{ (color online)
Magnetic wave propagation through GLHUA outer layer cloak, wave front at 24 step.
}\label{fig10}
\end{figure}
\begin{figure}[h]
\centering
\includegraphics[width=0.85\linewidth,draft=false]{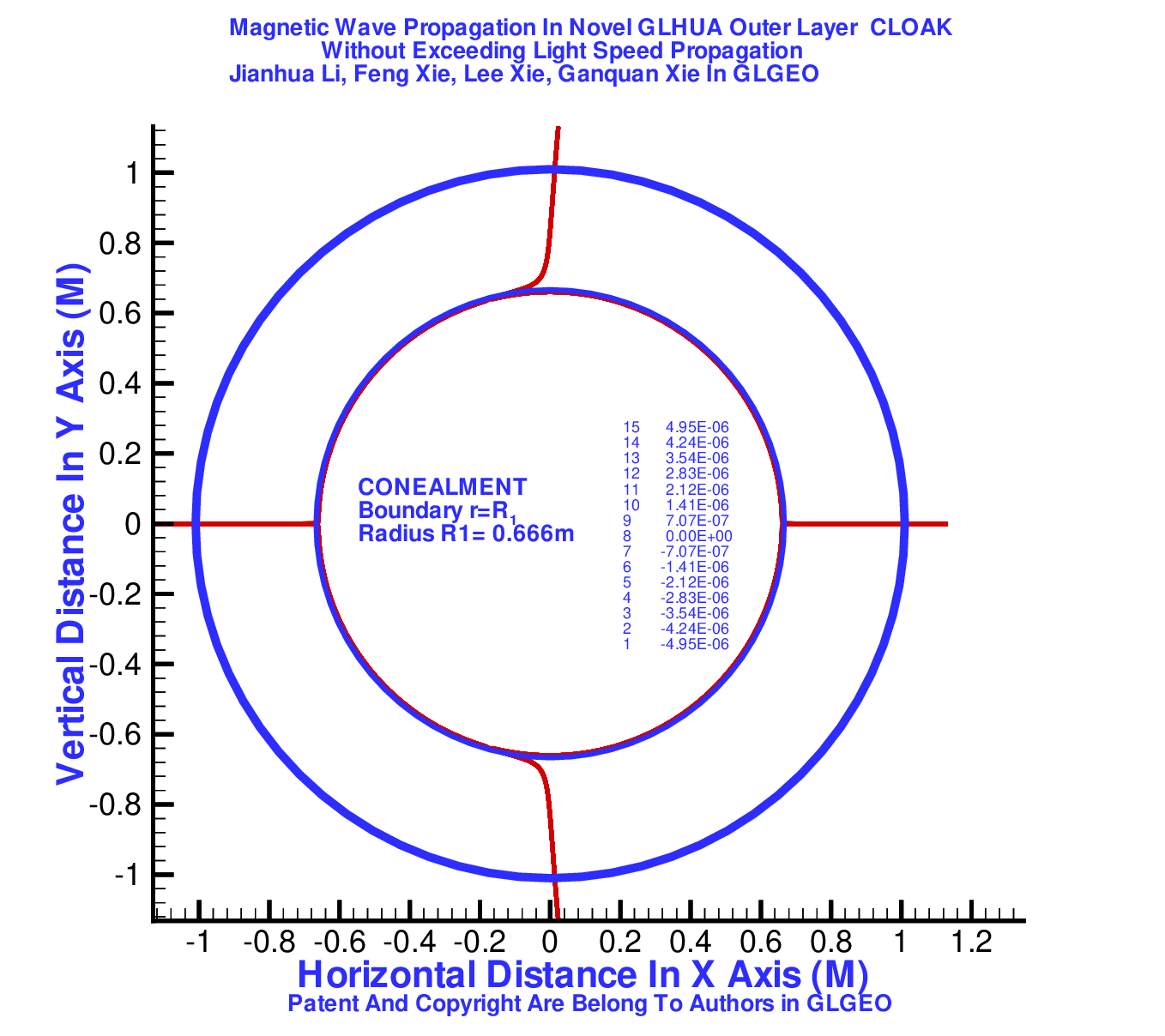}
\caption{ (color online) 
Magnetic wave propagation through GLHUA 
outer layer cloak, wave front at 25 step.
}\label{fig11}
\end{figure}
\begin{figure}[h]
\centering
\includegraphics[width=0.85\linewidth,draft=false]{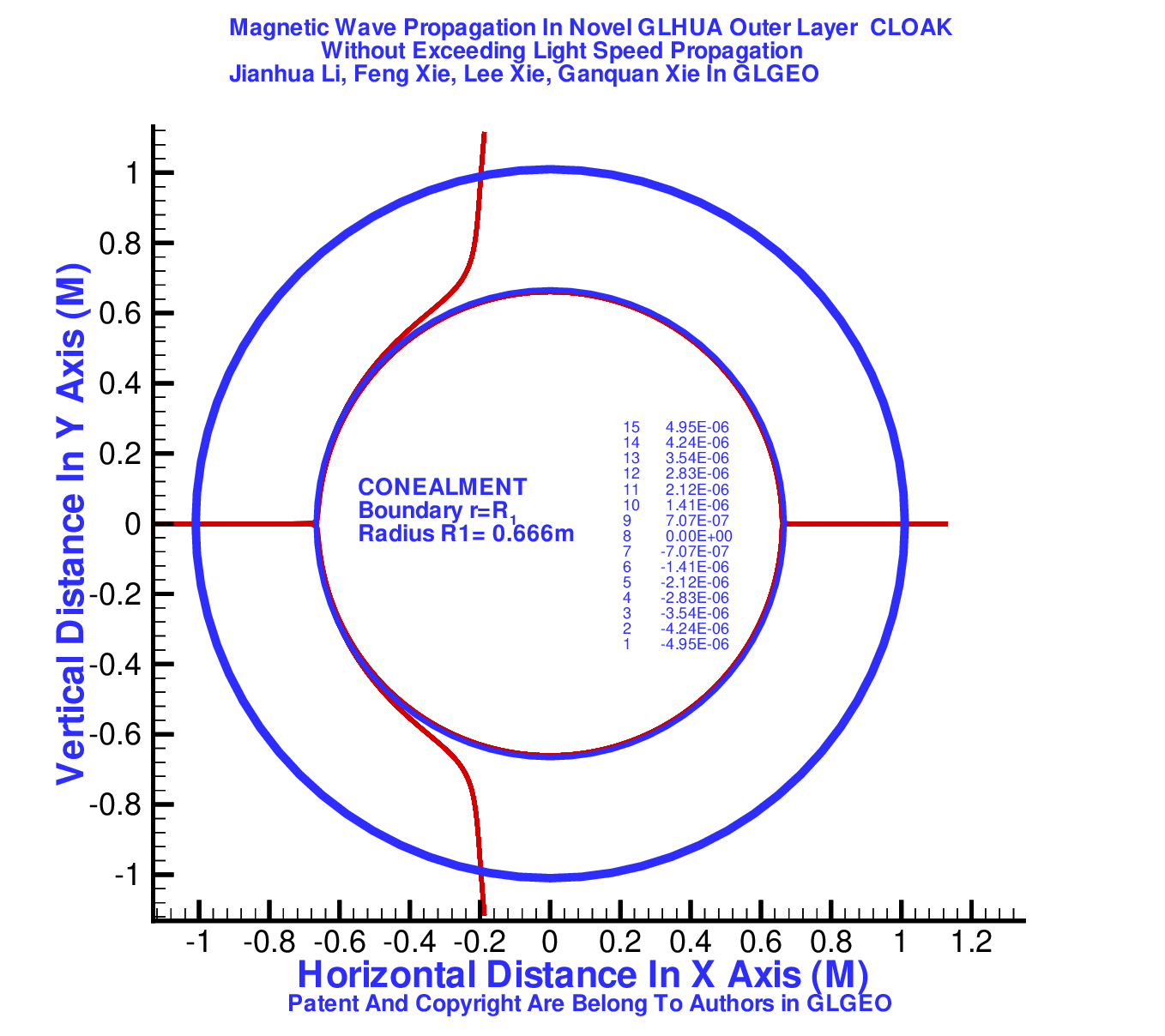}
\caption{ (color online) 
Magnetic wave propagation through GLHUA 
outer layer cloak, wave front at 29 step.
 }\label{fig12}
\end{figure}
\begin{figure}[h]
\centering
\includegraphics[width=0.85\linewidth,draft=false]{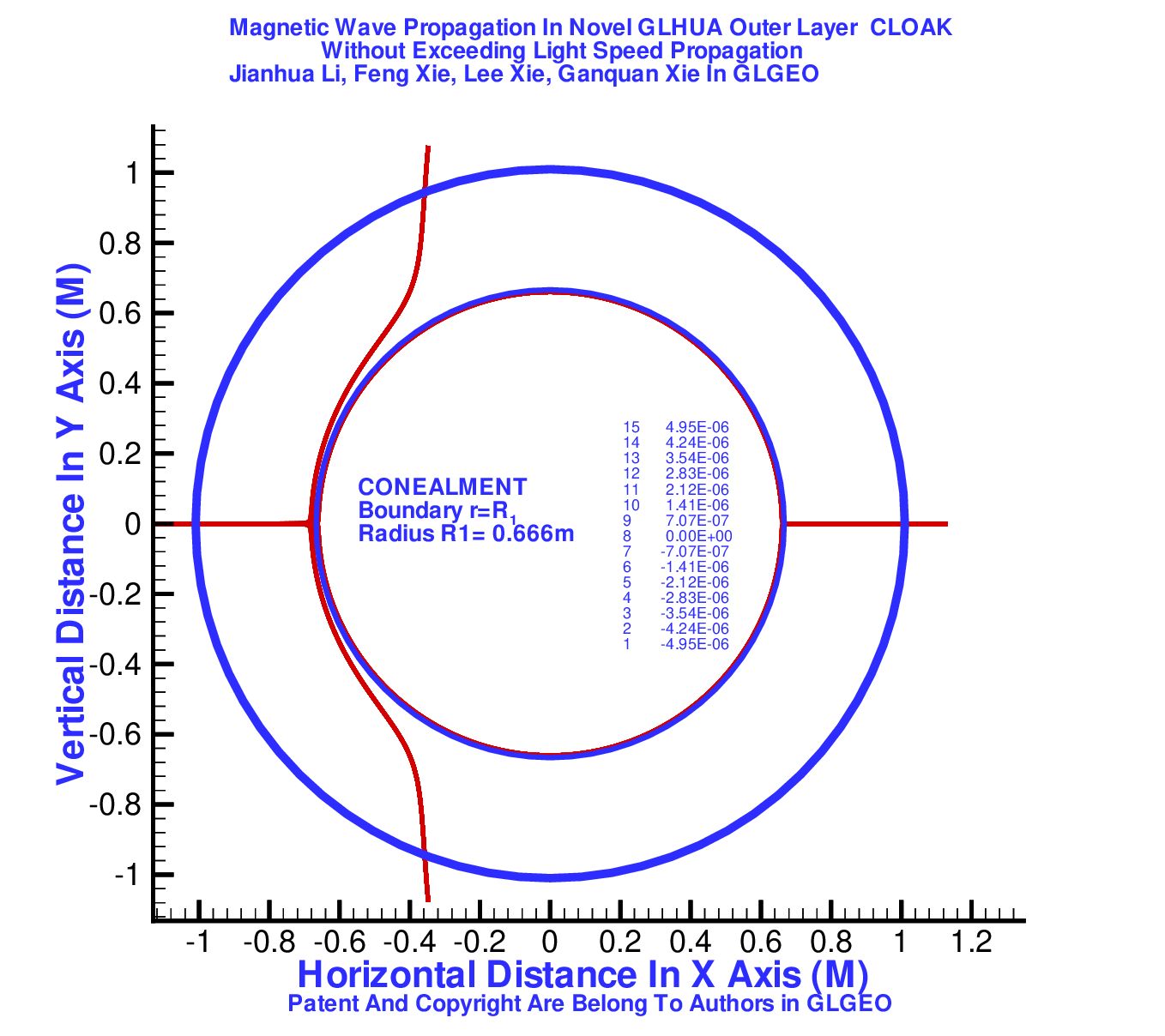}
\caption{ (color online) 
Magnetic wave propagation through GLHUA 
outer layer cloak, wave front at 32step.
}\label{fig13}
\end{figure}
\begin{figure}[h]
\centering
\includegraphics[width=0.85\linewidth,draft=false]{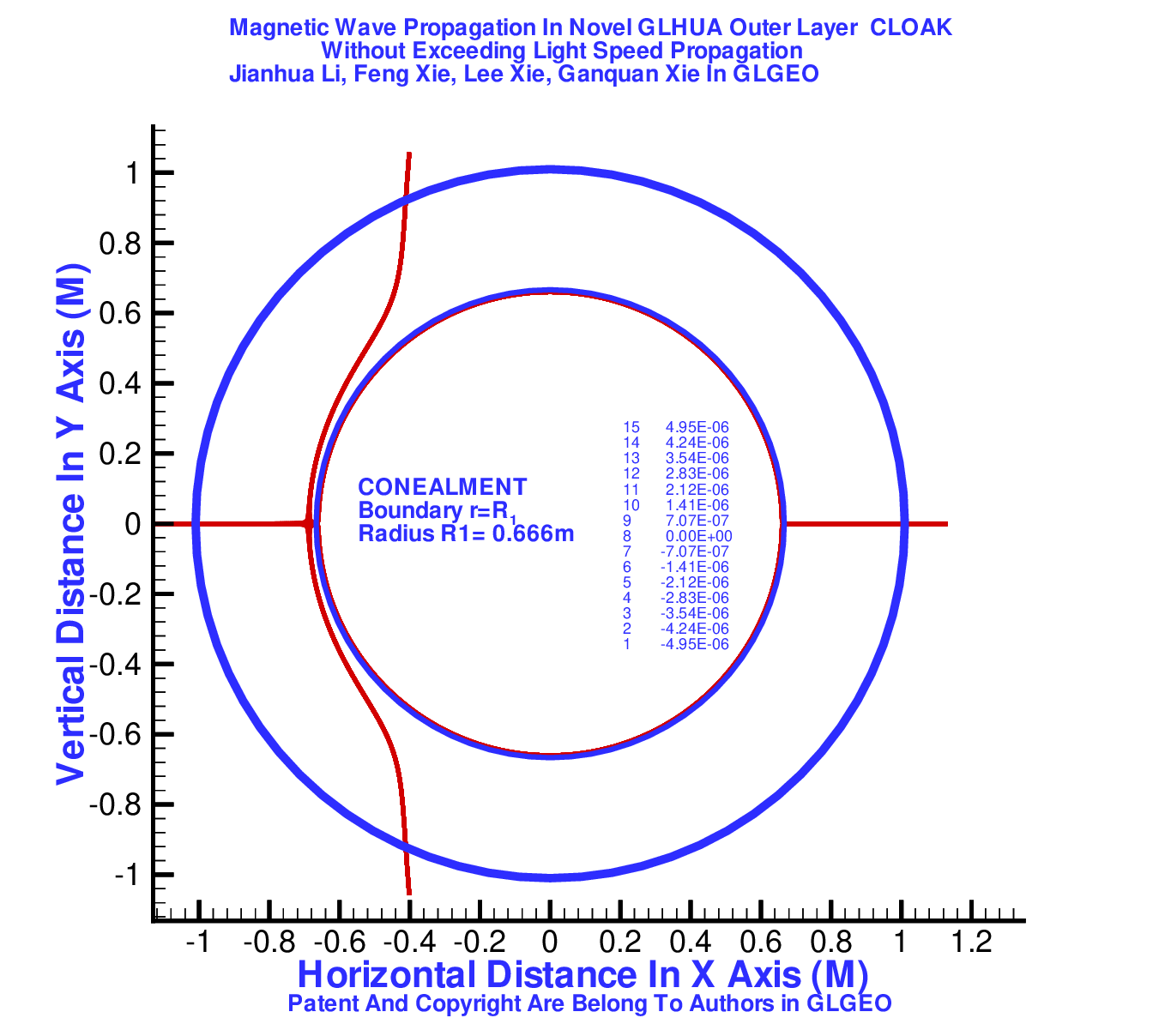}
\caption{ (color online) 
Magnetic wave propagation through GLHUA 
outer layer cloak, wave front at 33 step.
   }\label{fig14}
\end{figure}
\begin{figure}[h]
\centering
\includegraphics[width=0.85\linewidth,draft=false]{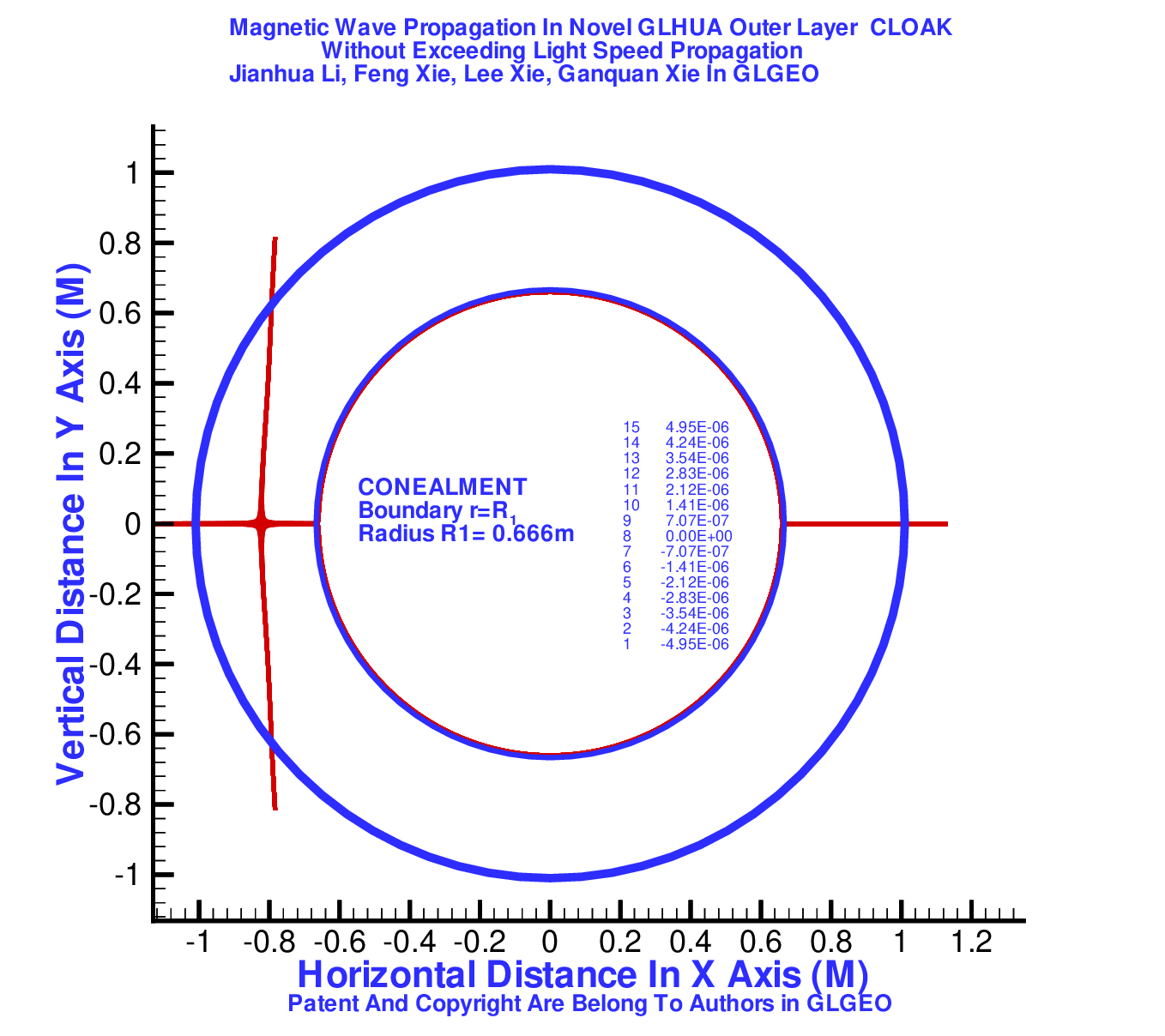}
\caption{ (color online) 
Magnetic wave propagation through GLHUA 
outer layer cloak, wave front at 40 step.
 }\label{fig15}
\end{figure}
\begin{figure}[h]
\centering
\includegraphics[width=0.86\linewidth,draft=false]{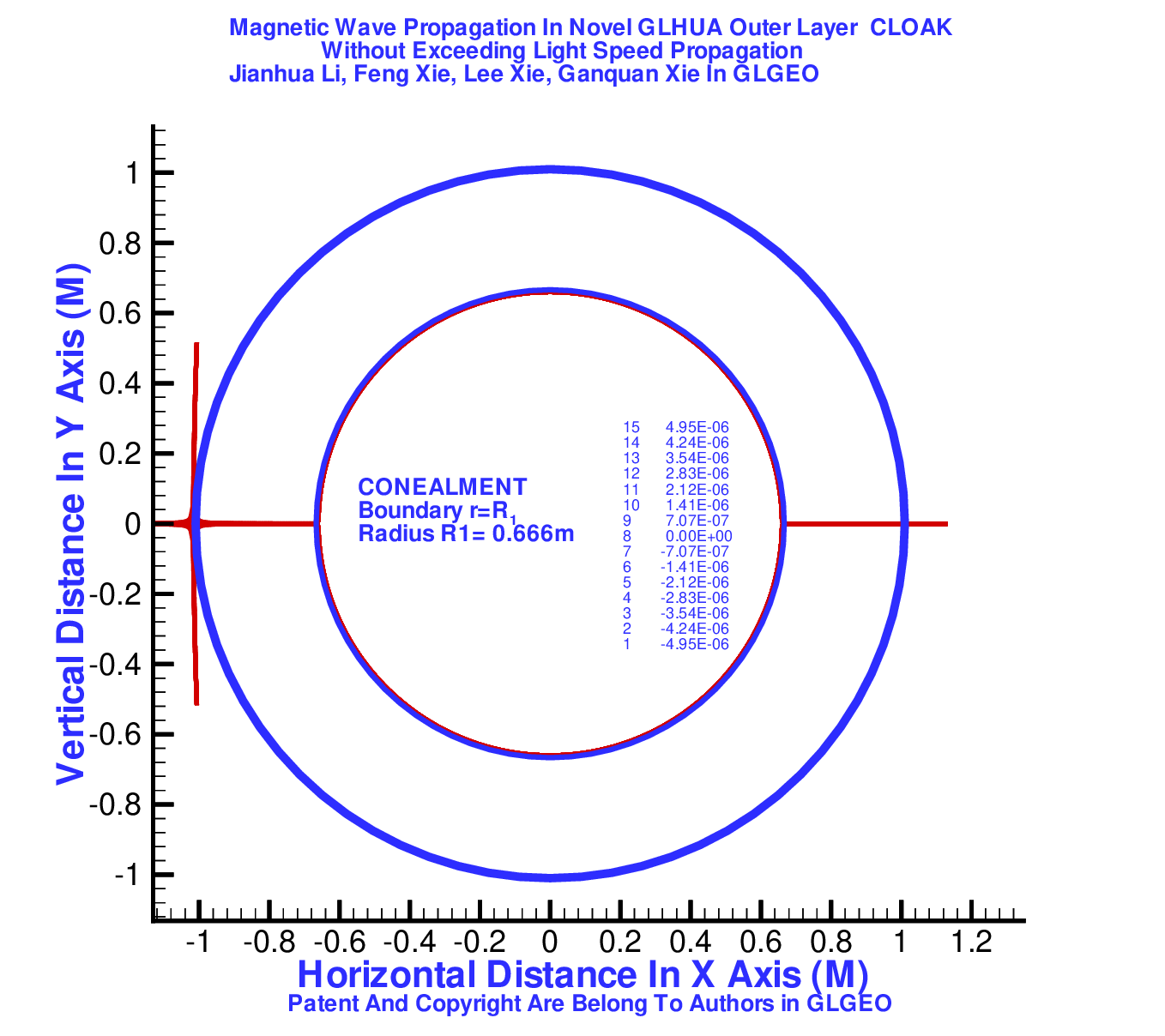}
\caption{ (color online) 
Magnetic wave propagation through GLHUA 
outer layer cloak, wave front at 44 step, Wave front is
recovered as the same as incident wave front.
 }\label{fig16}
\end{figure}
\begin{figure}[h]
\centering
\includegraphics[width=0.85\linewidth,draft=false]{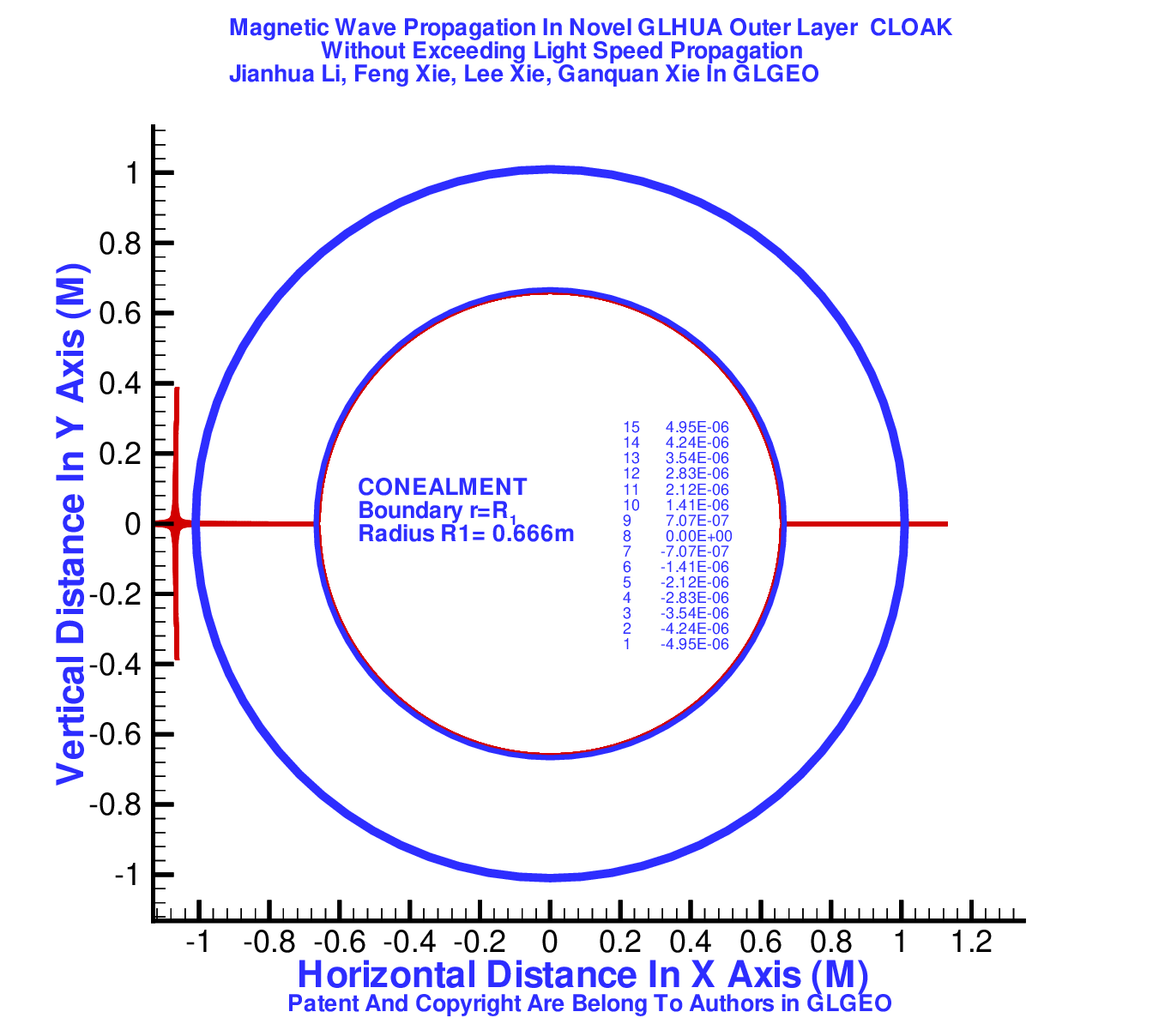}
\caption{ (color online) 
Magnetic wave propagation through GLHUA outer layer cloak, wave front at 45 step. Wave front is
recovered as the same as incident wave front.
}\label{fig17}
\end{figure}

\begin{figure}[h]
\centering
\includegraphics[width=0.86\linewidth,draft=false]{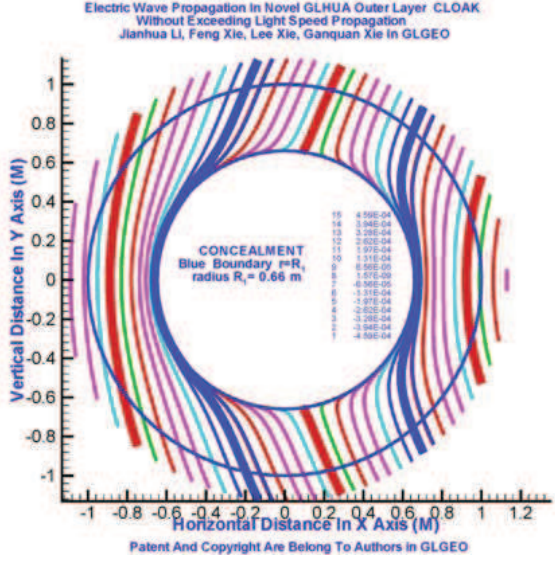}
\caption{ (color online) 
For frequency $f = {\rm 0}{\rm .16687773} \times 10^9 $ Hz, $rs=3m$, electric wave $Ex$ propagation through GLHUA cloak. The phase velocity without exceeding light speed and tends to zero in$ r=R1.$
 }\label{fig18}
\end{figure}
\begin{figure}[h]
\centering
\includegraphics[width=0.86\linewidth,draft=false]{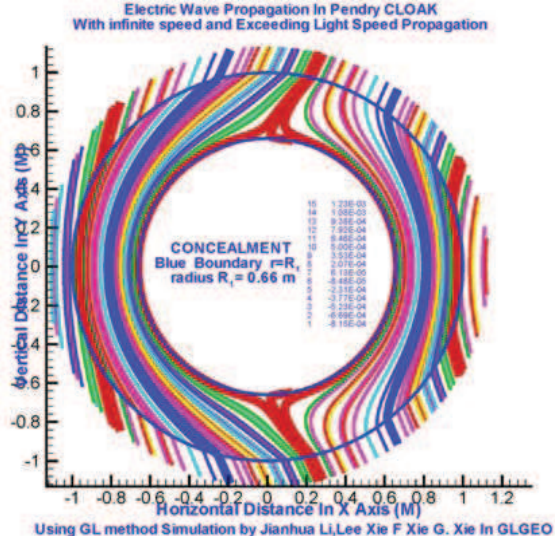}
\caption{ (color online) 
For frequency $f = {\rm 0}{\rm .16687773} \times 10^9 $ Hz rs=3m electric wave Ex propagation through Pendry cloak with infinite speed and with exceeding light speed.
 }\label{fig19}
\end{figure}

\begin{figure}[h]
\centering
\includegraphics[width=0.86\linewidth,draft=false]{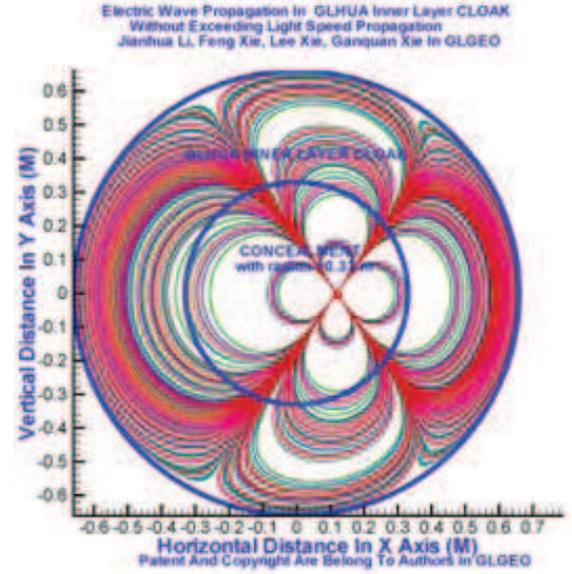}
\caption{ (color online) 
Electric wave Ex excited by electric source at (0.01833, 0, 0) in concealment propagation through GLHUA Inner layer cloak. The electric wave propagation can not arrive to r=R1, can not Propagation to outside of the inner layer. The electric wave in concealment can be not disturbed by the cloak. R1=0.66m, R0=0.33m.
}\label{fig31}
\end{figure}

\begin{figure}[h]
\centering
\includegraphics[width=0.86\linewidth,draft=false]{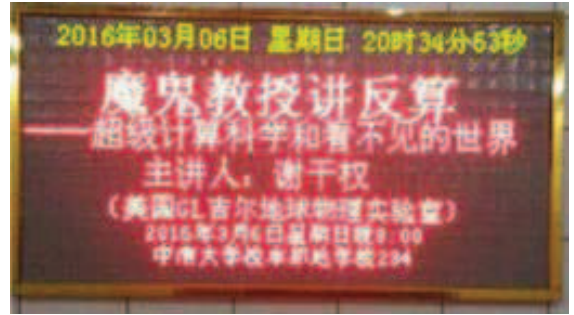}
\caption{ (color online) 
Chinese advertising by Central South University in Changsha on Msrch 6, 2016:
  Professor Ganquan Xie speak "Demon talk about inversion and invisible Super Sciences and
GLHUA practicable invisible cloak and 
Super Computational Sciences"}\label{fig32}
\end{figure}
\begin{figure}[h]
\centering
\includegraphics[width=0.86\linewidth,draft=false]{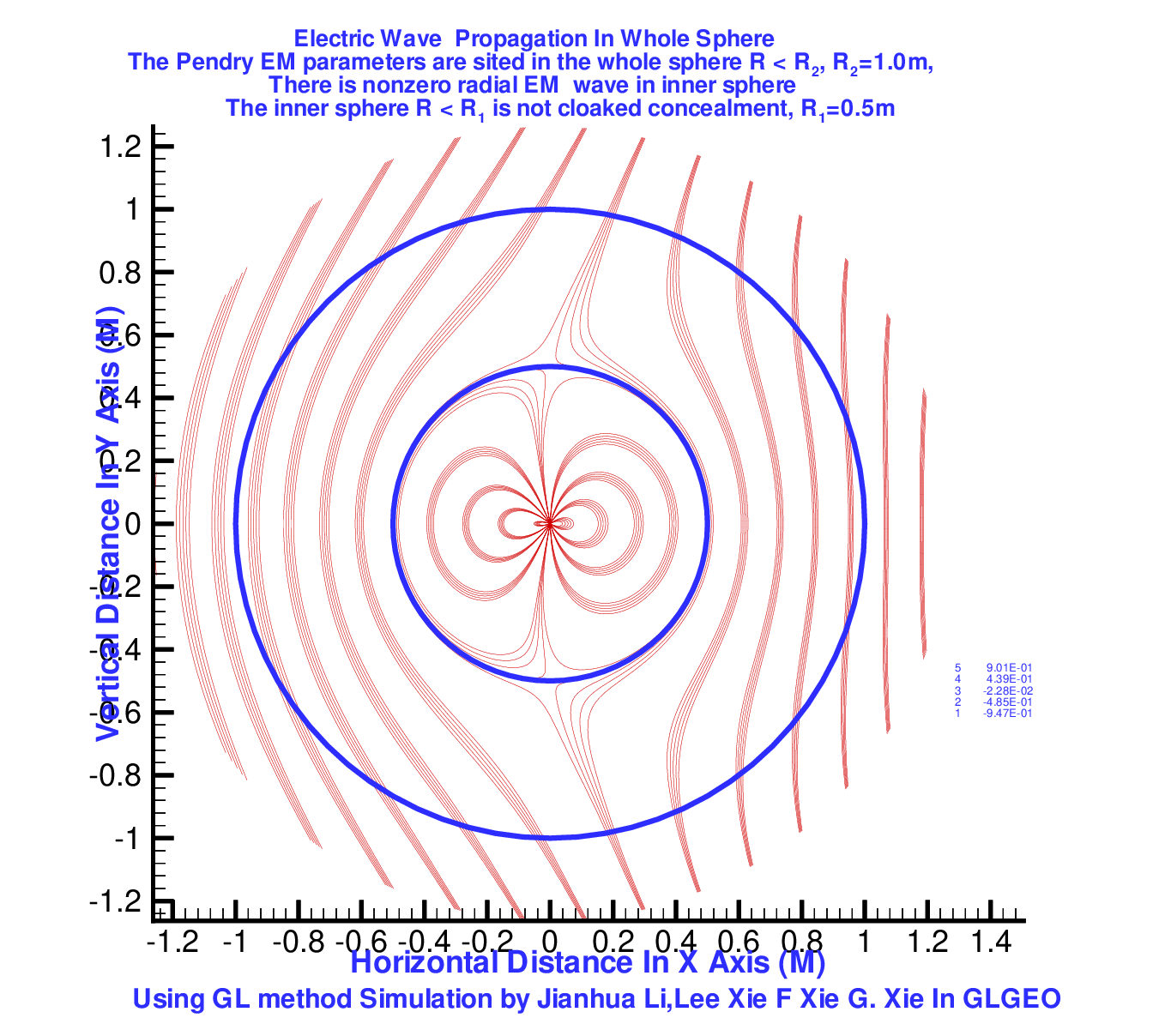}
\caption{ (color online) If Pendry EM parameters are sit in the whole sphere, $0 \le r \le R_2$   which include the annular layer  $R_1 \le r \le R_2$   and inner sphere $0 \le r \le R_1$ , then there is the
electromagnetic wave propagation through whole sphere $0 \le r \le R_2$     , there is nonzero
electromagnetic wave propagation in the inner sphere $0 \le r \le R_1$ , the inner sphere is not
cloaked concealment. 
}\label{fig33}
\end{figure}
\begin{figure}[h]
\centering
\includegraphics[width=0.86\linewidth,draft=false]{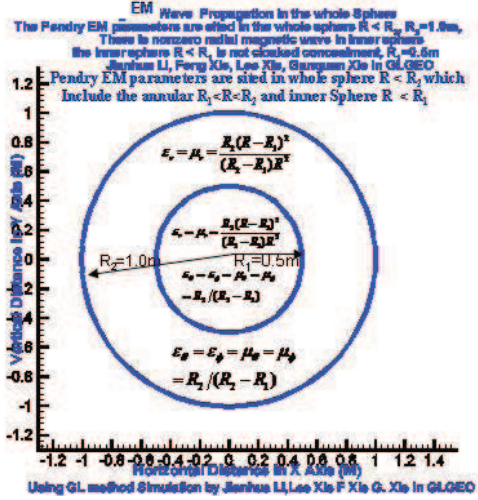}
\caption{ (color online)   Pendry EM parameters are sit in the whole sphere, $0 \le r \le R_2$   which include the annular layer  $R_1 \le r \le R_2$   and inner sphere $0 \le r \le R_1$. 
}\label{fig34}
\end{figure}

\section {GLHUA EM invisible double layers cloak materials and properties}
In this section, we propose GLHUA EM invisible double layers cloak materials by (1) and (2) ((1) and (12) in [1]); and the properties of GLHUA double layer complete invisible cloak. 
\subsection{GLHUA double layer EM cloak Materials with relative parameters that is not less than 1}
In the outer annular layer, $R_1  \le r \le R_2 $, the relative electric permittivity in (1) in [1]
\begin{equation}
\begin{array}{l}
 \varepsilon _r  = 1, \\ 
 \varepsilon _\theta   = \varepsilon _\phi   = \frac{1}{2}\left( {\left( {\frac{{r - R_1 }}{{R_2  - R_1 }}} \right)^\alpha   + \left( {\frac{{R_2  - R_1 }}{{r - R_1 }}} \right)^\alpha  } \right), \\ 
 \end{array}
\end{equation}
and the relative magnetic permeability, $\mu _r  = \varepsilon _r $,
$\mu _\theta   = \mu _\phi   = \varepsilon _\theta  $,
and $0 < \alpha _0  < \alpha  \le \alpha _1  < 2$.
In inner layer cloak $0 < R_0  \le r \le R_1 $, the relative electric permittivity in (2) (in (12) in [1]),
\begin{equation}
\begin{array}{l}
 \varepsilon _r  = 1, \\ 
 \varepsilon _\theta   = \varepsilon _\phi   = \frac{1}{2}\left( {\left( {\frac{{R_1  - r}}{{R_1  - R_0 }}\frac{{R_0 }}{r}} \right)^\alpha   + \left( {\frac{{R_1  - R_0 }}{{R_1  - r}}\frac{r}{{R_0 }}} \right)^\alpha  } \right), \\ 
 \end{array}
\end{equation}
the relative magnetic permeability, $\mu _r  = \varepsilon _r $,
$\mu _\theta   = \mu _\phi   = \varepsilon _\theta  $,
and $0 < \alpha _0  < \alpha  \le \alpha _1  < 2$.

\subsection{Properties of GLHUA Double Layer Electromagnetic Cloak}
${\boldsymbol{Property \ 1}}$,\ In outer layer of GLHUA double layer cloak, the relative electric permittivity and magnetic permeability parameter in (1) ((1) in [1]) and their derivative are continuous across boundary ,$r = R_2 $, outer boundary of outer annular layer of GLHUA double cloak. Proof is given in the thorem 2.1 in the section 2 of paper [1].
\hfill\break

${\boldsymbol{Property \ 2}}$,\ By GL full wave no scattering modeling and inversion theoretical analysis of solution of the Maxwell equation (1)-(4) in [1] and [29], in the outer annular layer cloak with GLHUA cloak material (1) ((1) in [1]), the EM wave excited in outside of cloak $r_s  > R_2 $ will be vanished at $r=R_1$, i.e. when r is decreasing and going to $R_1$, 
$\mathop {\lim }\limits_{r \to R_1 } \vec E(\vec r) = 0$
 and $\mathop {\lim }\limits_{r \to R_1 } \vec H(\vec r) = 0$.
Proof is given in the theorem 4.1 in section 4  of paper [1].

\hfill\break
 
${\boldsymbol{Property \ 3}}$,\ In the inner layer cloak of GLHUA double layer cloak, the relative electric permittivity and magnetic permeability parameter in (2) ((12) in [1]) and their derivative are continuous across boundary $r=R_0$, inner boundary of inner annular layer of GLHUA double cloak. Proof is given theorem 2.3 in the section 2 of paper [1].

\hfill\break
 
${\boldsymbol{Property \ 4}}$,\ By GL full wave no scattering theoretical analysis of solution of the Maxwell equation (1)-(4) in [29] and [1], in the inner annular layer cloak with GLHUA cloak material in (2) ((12) in [1]), the EM wave excited inside concealment,$r_s  < R_0 $  will be vanished at
$r=R_1$, i.e. when r is increase and going to $R_1$, $\mathop {\lim }\limits_{r \to R_1 } \vec E(\vec r) = 0$
 and $\mathop {\lim }\limits_{r \to R_1 } \vec H(\vec r) = 0$.
The proof is given in theorem in the section 6 of the paper [1].

From property 3 and property 4, the incident EM wave is excited inside of concealment of GLHUA double layer cloak, $r_s  < R_0 $, there is no scattering wave from the cloak to disturb the inside incident wave in concealment with free space material. The incident EM wave exited in outside of GLHUA cloak, $r_s  > R_2 $ , can not propagate enter the concealment sphere, $r < R_0 $  , The incident EM wave exited in inside of the concealment of GLHUA double layer cloak can not propagate to outside of GLHUA cloak. The EM wave in the outside of the GLHUA cloak can not be disturbed
the cloak, The EM wave in the outside of the GLHUA cloak equal to incident wave. The reciprocal principle is satisfied for GLHUA double layer cloak that proved in section 5 in [1].

\hfill\break
${\boldsymbol{Property \ 5}}$,\ The relative electric permittivity and magnetic permeability parameters are not less than 1, that proved in
theorem 2.2 and theorem 2.4 in section  2 in [1]             
\begin{equation}
\varepsilon _r  = \mu _r  \ge 1,\varepsilon _\theta   = \varepsilon _\phi   = \mu _\theta   = \mu _\phi   \ge 1,
\end{equation}

The property 5 shows that GLHUA double layer cloak with material parameter not less than 1 ( (1) for the outer layer and (12) for the inner layer in [1]), the EM wave propagation in GLHUA double layer cloak without exceeding light speed and without infinite speed propagation. GLHUA cloak can be practicable. GLHUA cloak overcome fundamental difficulties in Pendry cloak.

\section { Create GLHUA double layer cloak}

\subsection { Create relative radial parameter $\varepsilon _r  = \mu _r  = 1
$ in (1) ((1) in [1]) in GLHUA outer layer cloak $R_1  \le r \le R_2 $, and (2) ((12) in [1]) in inner layer $ R_0  \le r \le R_1 $.}

\subsection { Create the relative angular parameter
$\varepsilon _\theta   = \varepsilon _\phi   = \mu _\theta   = \mu _\phi   = \frac{1}{2}\left( {\left( {\frac{{r - R_1 }}{{R_2  - R_1 }}} \right)^\alpha   + \left( {\frac{{R_2  - R_1 }}{{r - R_1 }}} \right)^\alpha  } \right)$
in (1) in outer layer $R_1  \le r \le R_2 $, $0 < \alpha _0  < \alpha  \le \alpha _1  < 2$}
Because we did chose the radial relative parameter $\varepsilon _r  = \mu _r  = 1$ in the outer layer
cloak $R_1  \le r \le R_2 $, so we can not chose a finite relative angular parameter, otherwise the EM wave excited in the outside of the cloak should be propagation cross boundary $r=R_1$  and penetrate into the sphere $r<R_1$. Therefore, created the relative angular parameter should be going to infinite when $r \to R_1 $. Because radial GL electric equation (19) in [29] and magnetic wave equation (23) in [29] are of the same form, we should to create $\varepsilon _\theta  (r) = \mu _\theta  (r) = \varepsilon _\phi  (r) = \mu _\phi  (r)
$. For making parameter continuous and equal to 1 and its derivative continuous across boundary 
$r=R_2$ and equal to zero on $r=R_2$ , so, we create 

\[
\begin{array}{l}
 \varepsilon _\theta   = \varepsilon _\phi   = \mu _\theta   = \mu _\phi   = \frac{1}{2}\left( {\left( {\frac{{r - R_1 }}{{R_2  - R_1 }}} \right)^\alpha   + \left( {\frac{{R_2  - R_1 }}{{r - R_1 }}} \right)^\alpha  } \right), \\ 
 0 < \alpha _0  < \alpha  \le \alpha _1  < 2, \\ 
 \end{array}
\]

in (1) ((1) in [1]). The relative angular parameter is not unique. By GL modeling and no scattering inversion, the rigorous and detailed theoretical proof of GLHUA outer layer invisible cloak is given for $\alpha  = 1$ in the paper [1]. Similarly, we can prove GLHUA outer layer is invisible cloak for $0 < \alpha _0  < \alpha  < \alpha _1  < 2$.

\subsection{Create the relative angular parameter
$\varepsilon _\theta   = \varepsilon _\phi   = \mu _\theta   = \mu _\phi   = \frac{1}{2}\left( {\left( {\frac{{R_1  - r}}{{R_1  - R_0 }}\frac{{R_0 }}{r}} \right)^\alpha   + \left( {\frac{{R_1  - R_0 }}{{R_1  - r}}\frac{r}{{R_0 }}} \right)^\alpha  } \right)$
in (2) ((12) in [1]) in the inner layer cloak $R_0  \le r \le R_1 $, $0 < \alpha _0  < \alpha  \le \alpha _1  < 2$}

Because we did chose radial parameter $\varepsilon _r  = \mu _r  = 1$, in the inner layer cloak, so, we can not chose finite angular parameter, otherwise the EM wave excited in the inside of the concealment $r<R_0$ should be propagation cross boundary $r=R_1$ and go to the outside of the sphere $r>R_1$ . Therefore, created the relative angular parameter should be going to infinite when r is increasing and $r \to R_1$. Similarly, for making parameter continuous and equal to 1 and its derivative continuous across the boundary $r=R_0$ and equal to zero on the boundary $r=R_0$ , we create the angular parameters, 
\[
\begin{array}{l}
 \varepsilon _\theta   = \varepsilon _\phi   = \mu _\theta   = \mu _\phi   \\ 
  = \frac{1}{2}\left( {\left( {\frac{{R_1  - r}}{{R_1  - R_0 }}\frac{{R_0 }}{r}} \right)^\alpha   + \left( {\frac{{R_1  - R_0 }}{{R_1  - r}}\frac{r}{{R_0 }}} \right)^\alpha  } \right) \\ 
 \end{array}
\]
in (2) ((12) in [1]) in the inner layer cloak $R_0  \le r \le R_1 $.

\subsection{The theoretical analysis and proof of GLHUA double layer cloak by GL modeling and no scattering inversion}

Substitute GLHUA outer layer parameter (1), ((1) in [1]) into GL EM wave equation in the outer layer $R_1  \le r \le R_2 $, and interface conditions on boundary $r=R_1$ , the necessary condition to make that the wave can not penetrate into the sphere$ r< R_1$  is    
 
\begin{equation}
\vec E(R_1 ^ +  ,\theta ,\phi ) = 0,
\end{equation}

That is (40), (52) in [1], and

\begin{equation}
\vec H(R_1 ^ +  ,\theta ,\phi ) = 0,
\end{equation}

That is (41), (53) in [1] and
 
\begin{equation}
\frac{1}{{\varepsilon _\theta  (R_1 ^ +  )}}\frac{\partial }{{\partial r}}\vec E(R_1 ^ +  ,\theta ,\phi ) = 0.
\end{equation}

That is (48), (71) in [1], and

\begin{equation}
\frac{1}{{\mu _\theta  (R_1 ^ +  )}}\frac{\partial }{{\partial r}}\vec H(R_1 ^ +  ,\theta ,\phi ) = 0.
\end{equation}

That is (49), (72) in [1]. Where  $R_1 ^ + $ means limitation when $ r \ge R_1 $ and $ r \to R_1 $ . The above conditions are proved in the theorem 4.1-4.4 in [1]. Similarly, the necessary condition to make that the EM wave excited in concealment sphere $ r < R_0  $ can not be propagated to outer $r > R_1 $ is that               
\begin{equation}
\vec E(R_1^ -  ,\theta ,\phi ) = 0,
\end{equation}

\begin{equation}
\vec H(R_1^ -  ,\theta ,\phi ) = 0,
\end{equation} 

and

\begin{equation}
\frac{1}{{\varepsilon _\theta  (R_1 ^ -  )}}\frac{\partial }{{\partial r}}\vec E(R_1 ^ -  ,\theta ,\phi ) = 0.
\end{equation}

\begin{equation}
\frac{1}{{\mu _\theta  (R_1 ^ -  )}}\frac{\partial }{{\partial r}}\vec H(R_1 ^ -  ,\theta ,\phi ) = 0.
\end{equation}

where $R_1 ^ -  $ means limitation when  $r \le R_1 $ and $r \to R_1 $. These conditions can be proved similar with theorem 4.1-4.4 in [1]. Summary due to GLHUA outer layer parameter (1) ((1) in [1]) and its derivative are continuous across the boundary$r= R_2$, and conditions (4),(6);(5),(7) on the boundary $ r=R_1$ are satisfied, in the theorem 5.1-5.3 in [1], we proved that the incident EM wave excited in the outside of GLHUA cloak can not penetrate into the sphere $ r < R_1$. In the theorem 5.4-5.6 in [1], we proved that the incident EM wave excited in the outside of GLHUA cloak can not be disturbed by the cloak. Similarly, due to GLHUA inner layer parameter (2) ((12) in [1]) and its derivative are continuous across the boundary $ r= R_0$, and boundary conditions (8) (10); (9) (11), on the boundary $ r= R_1$ are satisfied. In the theorem 6.1-6.3 in [1], we proved that the incident EM wave excited in the inside of the concealment  $r < R_0 $ can not propagation to go to out to $ r> R_1$. In the theorem 6.4-6.6 of in [1], we proved that the incident EM wave excited in the inside of the concealment $ r < R_0$ can not be disturbed by the cloak. Therefore, GLHUA double layer cloak is invisible cloak with the sphere concealment $r < R_0$. Because GLHUA double layer cloak with relative parameter not less than 1, then the EM wave propagation through GLHUA cloak without infinite speed and without exceeding light speed. The Zero condition of wave on the boundary $R_1^ +  $, i.e. the outer side of the inner boundary of GLHUA outer layer cloak, (4),(6);(5),(7) is defined and proved in (40) (41); (52),(53) in [1]. Similarly, we defined the zero conditions (8) (10); (9) (11), of wave on $R_1^ -  $, i.e., the inner side of the outer boundary of the inner layer cloak.

\subsection{ Proof of zero condition of wave on the outer side of inner boundary $R_1^ +  $ of GLHUA outer layer cloak}

Using GL no scattering modeling inversion with generalized regularization [2]
[3][24][25], $\vec E(R_1^ +  ,\theta ,\phi ) = 0$, in (4) is in (40), (52) in [1], and $\vec H(R_1^ +  ,\theta ,\phi ) = 0$, in (5) is in (41), (53) in [1] in generalized regularization that are proved in the theorem 4.1 and the theorem 4.3 respectively in [1]. $\frac{1}{{\varepsilon _\theta  (R_1^ +  )}}\frac{\partial }{{\partial r}}\vec E(R_1^ +  ,\theta ,\phi ) = 0$, in (6) (in (48) in [1]), and $\frac{1}{{\mu _\theta  (R_1^ +  )}}\frac{\partial }{{\partial r}}\vec H(R_1^ +  ,\theta ,\phi ) = 0,$ in (7) (in (49) in [1]) are proved in the theorem 4.2 and the theorem 4.4 respectively in [1]. The proof of the theorem 4.1-4.4 in the section 4 in [1] is based on the theorem 6.1-6.4 in the section 6 in [29]. The proof of theorem 6.1-6.4 is based on GLHUA pre cloak material conditions (6.1)-(6.4) and GL EM wave field equation in the virtual invisible sphere in [29]. From GLHUA pre cloak angular parameter condition (6.4) in GLHUA invisible virtual sphere, a novel transform between the physical GLHUA cloak outer annular layer and the virtual invisible sphere is derived in (25) in [1], GL radial transform is 
\begin{equation}
r_p  = R_1  + p(r) = R_1  + Ae^{ - \frac{B}{r}} ,
\end {equation}
that is (25) in the section 3 in [1]. Also, from GLHUA pre cloak angular parameter condition (6.4), we derived the relative angular parameter in (38) in [1] that is the same as (1) ((1) in [1]). The proof of theorems 6.1 - 6.4 is given in [29]. In the section 2 in [29], we proposed GL radial the EM wave field $E(\vec r)$ and $H(\vec r)$  in (18), GL electric wave equation (19), and GL magnetic wave equation (23) in [29]. Equations (19) and (23) in [29] are of the same form and self adjoint equations. We did find and proved essential property of GL radial EM wave that the sphere surface integral of $E(\vec r)\sin \theta $ and $H(\vec r)\sin \theta $ are zero in theorem 3.1-3.2
in [29] that is an essential property of EM wave. That is important for proof of the theorems 6.1-6.4 in [29]. Similarly, we can prove zero conditions of the wave on $R_1^ -$, the inner side of the outer boundary of GLHUA inner layer cloak. Because the relative angular parameter is denoted by (1) ((1) in [1]), using $\vec E(R_1^ +  ,\theta ,\phi ) = 0$  and $\vec H(R_1^ +  ,\theta ,\phi ) = 0$   the following conditions can be respectively proved:

\begin{equation}
\frac{1}{{\varepsilon _\theta  (R_1 ^ +  )}}\frac{\partial }{{\partial r}}\vec E(R_1 ^ +  ,\theta ,\phi ) = 0,
\end{equation}

and

\begin{equation}
\frac{1}{{\mu _\theta  (R_1 ^ +  )}}\frac{\partial }{{\partial r}}\vec H(R_1 ^ +  ,\theta ,\phi ) = 0.
\end{equation}

$\vec E(R_1^ +  ,\theta ,\phi ) = 0$ in (4) (in (40), (52) in [1]), and $\vec H(R_1^ +  ,\theta ,\phi ) = 0$ in (5) (in (41),(53) in [1]) are proved in theorem 4.1 and theorem 4.3 respectively in [1]. That is other approach to prove zero condition of the wave on $R_1^ +$, outer side of the inner boundary of GLHUA outer layer cloak. That is the merit of the relative angular parameter (1) ((1) in [1]) in GLHUA outer layer cloak.

\section{ GL no scattering modeling method is used for simulation of the EM wave propagation through GLHUA outer layer Cloak.}

In GLHUA outer layer cloak $R_1  < r \le R_2 $ with free space concealment $r < R_1$, we use GL EM full wave field no scattering modeling method [5-6][15] to simulate the EM wave propagation through GLHUA cloak, the wave is excited in the outside of the cloak, the source location in $\vec r_s  = (3.0m,0.5\pi ,0)$. The radius of the outer boundary in the outer sphere annular layer is 1.0 meter, $R_2  = 1.0m$, the radius of the inner boundary is 0.66 meter, $R_1  = 0.66m$, the electric current point source in direction $\vec e_x$.
\begin{equation}
S(\vec r,\vec r_s ) = e^{i\omega t} \delta (\vec r - \vec r_s )\vec e,
\end{equation}
         
in the free space $r \ge R_2$,$r \ge R_2$, $\omega  = 2\pi f$, the incident electric wave $E_x ^b $, incident magnetic wave $H_x ^b $. The EM wave propagation images are plotted in XZ plane. For the incident electric wave,   

\begin{equation}       
 E_x ^b  = e^{i\omega t} \left( {\frac{{\partial ^2 g}}{{\partial x^2 }} + k^2 g} \right){\raise0.7ex\hbox{${}$} \!\mathord{\left/
 {\vphantom {{} {i\omega \varepsilon _0 }}}\right.\kern-\nulldelimiterspace}
\!\lower0.7ex\hbox{${i\omega \varepsilon _0 }$}}
\end{equation}

The imaging of electric wave propagation through GLHUA cloak plotted in Figure 15 with frequency, $f = {\rm 0}{\rm .16687773} \times 10^9 $ Hz, for the same incident wave, the imaging of electric wave propagation through Pendry cloak plotted in Figure 16. By comparison the figure 15 and 16,
 GL simulation imaging show that the phase velocity of the EM wave propagation through GLHUA cloak is without exceeding light speed and tends to zero at boundary $r =R_1 $. However, the phase velocity of the EM wave propagation through Pendry cloak is with exceeding light speed and tends to infinity at boundary $r =R_1 $. Figure 2 to Figure 14 show that the magnetic wave  $H_y (\vec r)$ propagation through GLHUA cloak in time step by step. The incident magnetic wave, $H_y ^b  = \frac{1}{{ - i\omega \mu _0 }}e^{i\omega t} \frac{{\partial g}}{{\partial z}},$ in (17) is excited by electric point source with the frequency, $f = {\rm 0}{\rm .3347773} \times 10^8 $, the source location $\vec r_s  = (13.0m,0.5\pi ,0),$. In Figure 2, the incident magnetic wave propagation in the outside of the cloak and is not disturbed by the cloak. In Figure 3, the magnetic wave is being tangent to the outer boundary $r=R_2$ without any scattering. In the next step, the magnetic wave smoothly enter the outer annular layer, $R_1 < r <R_2$ ,is presented in Figure 4. In the Figure 5, the wave is starting to tangent to the inner boundary $r=R_1$. In next, Figure 5 to Figure 7 show that the magnetic wave front is splitting to the upper wave front and the lower wave front branches and their propagation. When the magnetic wave front tangentially contact with the inner spherical surface boundary $r=R_1$, at the contact point in Figure 5, the magnetic wave front is split into upper wave front and lower wave front branches. The Figures from 6 to 7 show that the upper wave front is climbing up along the right upper hemispherical to Arctic. Figure 8 shows the upper wave front change propagation and start to slide down on Arctic. Figures 8 to 10 show that the upper magnetic wave front sliding down along the left upper hemispherical to Equatorial. In the meantime, the Figures 6 to 7 show that the lower wave front is sliding down along the right lower hemispherical to Antarctic, the Figure 7 shows that
 the wave front change propagation and start to climb up on Antarctic. Figures 8 to 10 show that the lower magnetic wave front climbing up along the left lower hemispherical to Equatorial. In intersection between Equatorial and the inner boundary $r=R_1$, the upper wave front and the lower wave front is merging into complete wave front in Figure 10. The EM wave front is propagation around the inner spherical boundary, $r=R_1$, the wave can not penetrate into the concealment. Figure 11 to Figure 13 show that completed magnetic wave front is continuous propagation to left without infinite speed and without exceeding light speed. When the wave front is leaving GLHUA cloak, Figure 13 shows that the magnetic wave front is the same as the incident wave front. Therefore, there is no any scattering from GLHUA cloak to disturb the incident wave in Figure 14 in free space; the wave can not penetrate into the concealment. GL simulation shows that GLHUA cloak is a completely invisible cloak with concealment without infinite speed and without exceeding light speed propagation. The comparison between Figure 15 and Figure 16 and the magnetic wave propagation from the figure 2 to the figure 14 show that the phase velocity of EM propagation through GLHUA cloak is no exceeding light speed and tends to zero in $r=R_1$, but the phase velocity of EM wave through Pendry cloak is exceeding light speed and tends infinity in $r=R_1$. Figure 17 shows that electric wave Ex excited by electric point source at (0.01833,0,0) in the concealment propagation through GLHUA Inner layer cloak with $R_1=0.66m$. GL method simulation shows that the EM wave propagation excited in the concealment can not arrive to $r=R_1$, and can not propagation to outside of the inner layer and can not propagation to outside of whole cloak. Moreover, the EM wave in the concealment can not be disturbed by the cloak.
In paper by Pendry et al in Science in 2006 [4], Pendry did state that nor can any radiation get out. that is totally wrong. We in [8][15] and Zhang and Chen in [7] proved that any radiation excited in Pendry cloak concealment will be propagation go to out. The reciprocal principle is not satisfied for Pendry cloak and is not satisfied for any single layer cloak. For overcoming the difficulty, we proposed inner layer cloak in GLHUA double layer cloak in this paper and in [9] in 2009. Up to now, only our inner cloak of the double layer cloak is discovered and proposed. The EM wave excited in the concealment $r < R_0$ can not be disturbed by cloak. The wave propagation can be not arrive $r=R_1$ and can be not go out of inner layer cloak. The wave propagation in GLHUA inner layer cloak is shown in Figure 17. The double layer 
$[R_0, R_1]$ and $[R_1, R_2]$  in GLHUA double layer cloak can be splitting into the inner layer $[R_0, R_1]$ with parameter (2) ((12) in [1]) and the outer layer $[R_2, R_3]$  with parameter
\begin{equation}
\begin{array}{l}
 \varepsilon _r  = 1, \\ 
 \varepsilon _\theta   = \varepsilon _\phi   = \frac{1}{2}\left( {\left( {\frac{{r - R_2 }}{{R_3  - R_2 }}} \right)^\alpha   + \left( {\frac{{R_3  - R_2 }}{{r - R_2 }}} \right)^\alpha  } \right), \\ 
 \end{array}
\end{equation}
in (1) ((1)in [1]), using $R_2$ to replace $R_1$ and $R_3$ to replace $R_2$, the middle layer 
$[R_1, R_2]$ can be free space or water to adjust temperature in practice.
GLHUA cloak and GLLH cloak published in paper arXiv1005.3999 are different class. Some numerical dispersion in GL method simulation of GLLH cloak in the paper arXiv1005.3999 [11] has been improved.  
Chen at al proved Pendry Cloak is invisible by Mei transform method [28], which is only available for constant relative angular parameter. 
Using GL no scattering modeling and inversion, we easy to prove that in Pendry cloak, 

\begin{equation}
\begin{array}{l}
 \mathop {\lim }\limits_{r \to R_1 } E\left( {\vec r} \right) = r^2 \varepsilon _r E_r  = 0, \\ 
 \mathop {\lim }\limits_{r \to R_1 } H\left( {\vec r} \right) = r^2 \mu _r H_r \left( {\vec r} \right) = 0, \\ 
 \end{array}
\end{equation}

\begin{equation}
\begin{array}{l}
 \mathop {\lim }\limits_{r \to R_1 } \frac{1}{{\varepsilon _\theta  }}\frac{\partial }{{\partial r}}E\left( {\vec r} \right) =  \\ 
  = \mathop {\lim }\limits_{r \to R_1 } \frac{1}{{\varepsilon _\theta  }}\frac{\partial }{{\partial r}}\left( {r^2 \varepsilon _r E_r \left( {\vec r} \right)} \right) = 0 \\ 
 \end{array},               
\end{equation}
\begin{equation}
\begin{array}{l}
 \mathop {\lim }\limits_{r \to R_1 } \frac{1}{{\mu _\theta  }}\frac{\partial }{{\partial r}}H\left( {\vec r} \right) =  \\ 
  = \mathop {\lim }\limits_{r \to R_1 } \frac{1}{{\mu _\theta  }}\frac{\partial }{{\partial r}}\left( {r^2 \mu _r H_r \left( {\vec r} \right)} \right) = 0, \\ 
 \end{array}.                 
\end{equation}
By theorem 5.1-5.6 in section 5 of the paper [1], we proved that Pendry Cloak is invisible cloaking for incident wave excited in outside of the cloak. However, any incident wave excited in the concealment will propagate to outside of the cloak. The reciprocal principle is not satisfied in Pendry cloak. 

\section{Comparison between GL full wave no scattering modeling method and ray tracing method for cloak}
\subsection {The head terms of the radial electric and magnetic wave equation (5),(6) in [31] and acoustic equation (1 ) in [32] in sphere coordinate are different type operator}

\subsubsection {The head terms of the radial electric and magnetic wave equation (5),(6) in [31] and acoustic equation (1 ) in [32] in sphere coordinate are different type operator} 

Compare the radial electric wave equation (5),magnetic wave equation (6) in [31] in free space and acoustic wave equation (1) in [32], their head terms are different type operator:
The head term of the radial electric wave equation (5) in [31] is
\begin{equation}
\frac{\partial }{{\partial r}}\frac{\partial }{{\partial r}}r^2 E_r \end {equation}                                                       ,                                                             
The head term of the radial magnetic wave equation (6)in [31] in free space is
 \begin{equation}
\frac{\partial }{{\partial r}}\frac{\partial }{{\partial r}}r^2 H_r 
\end {equation}                                                         
The head term of the acoustic wave equation (1) in [32] is
 \begin{equation}
\frac{\partial }{{\partial r}}r^2 \frac{\partial }{{\partial r}}u
\end{equation}                                                        ,                                                              
It is obvious that the head terms  in (21) and  in (22) are same type operator
\begin{equation} 
\frac{\partial }{{\partial r}}\frac{\partial }{{\partial r}}r^2 (V)
\end{equation}                                                             
where $V$ denotes the radial electric field $E_r$ in (21) or the magnetic field $H_r$  in (22).  However, the head term,$\frac{\partial }{{\partial r}}r^2 \frac{\partial }{{\partial r}}u$  in (23) of the acoustic wave equation (1)in [32]  has other different type operator
\[
\frac{\partial }{{\partial r}}r^2 \frac{\partial }{{\partial r}}u,  \ \  (23)
\]                                                                                                                      The head operator,$\frac{\partial }{{\partial r}}\frac{\partial }{{\partial r}}r^2 (V)
$  in (24) of the radial electric (5) and magnetic equation (6) in [31] is different from the head operator ,$\frac{\partial }{{\partial r}}r^2 \frac{\partial }{{\partial r}}u$  in (23)    of the acoustic equation (1)
in [32].
\
\subsubsection{The ray tracing method can not identify the different head terms between the radial EM equations and the acoustic equation}

In the almost all publications. the general ray tracing method can not identify the different head terms between the radial EM equations and the acoustic equation. 
The different head terms are basic different property between the radial EM equations and the acoustic equation that is key important to study cloak. Therefore. only ray tracing method is not theoretical method for studying cloak. GL full wave no scattering
modeling is important for studying cloak.
 \ 
\subsection { 0 in the BS to $R_1$   in the PS and $R_2$   invariant monotone increase differentiable radial transformation can be used to create electromagnetic invisible cloak}

Let 
\begin{equation}
r = R_1  + Q(R),
\end{equation}
be mapping $0$  in the BS   to  $R_1$  in the PS   and $R_2$   invariant monotone increase differentiable radial transformation.
Because  
\begin{equation}
\begin{array}{l}
 \mathop {\lim }\limits_{R \to 0} R^2 E_R  = 0, \\ 
 \mathop {\lim }\limits_{R \to 0} \frac{\partial }{{\partial R}}R^2 E_R  = 0, \\ 
 \end{array}
\end{equation}
 under the transformation (25),we have
 \begin{equation}
r^2 \varepsilon _r E_r  = R^2 E_R ,
\end {equation}                                     
                               
\begin{equation}
\mathop {\lim }\limits_{r \to R_1 } r^2 \varepsilon _r E_r  = \mathop {\lim }\limits_{R \to 0} R^2 E_R  = 0,
\end{equation} 
\begin{equation}
\frac{1}{{\mu _{\theta}   (r)}}\frac{\partial }{{\partial r}}r^2 \varepsilon _r E_r  = \frac{\partial }{{\partial R}}R^2 E_R ,
\end{equation}                              ,                                         
 
\begin{equation}
\mathop {\lim }\limits_{r \to R_1 } \frac{1}{{\mu _{\theta}   (r)}}\frac{\partial }{{\partial r}}r^2 \varepsilon _r E_r  = \mathop {\lim }\limits_{R \to 0} \frac{\partial }{{\partial R}}R^2 E_R  = 0
\end{equation} 

Similarly, we can prove that
\begin{equation}
\begin{array}{l}
 \mathop {\lim }\limits_{r \to R_1 } r^2 \mu _r H_r  = \mathop {\lim }\limits_{R \to 0} R^2 H_R  = 0, \\ 
 \mathop {\lim }\limits_{r \to R_1 } \frac{1}{{\varepsilon _\theta  (r)}}\frac{\partial }{{\partial r}}r^2 \mu _r H_r  = \mathop {\lim }\limits_{R \to 0} \frac{\partial }{{\partial R}}R^2 H_R  = 0, \\ 
 \end{array}
\end{equation}                          
Therefore, $0$  in the BS   to  $R_1$   in the PS   and $R_2$ invariant monotone increase differentiable radial transformation can be used to create electromagnetic invisible cloak.[4][28], In [28], authors used Mie transformation to prove Pendry cloak [4] is invisible cloak, but the Mie
transformation only suitable for constant relative angular EM parameters,
for example, for Pendry cloak case. 
but can not be for proving non constant angular EM parameters case.
Our GL method can be useful to prove that any $0$  in the BS   to  $R_1$   in the PS   and $R_2$ invariant monotone increase differentiable radial transformation can be used to induce EM invisible cloak.However,this type invisible cloak including Pendry cloak are impracticable cloak with infinite phase velocity at $r=R_1$ and with
exceeding light speed propagation physical difficulties.

\subsection{$0$ in the BS   to  $R_1$ in the PS   and $R_2$ invariant monotone increase differentiable radial 
transformation can not be used to create acoustic no scattering cloak}

Under the transformation (25), 
 \begin{equation}
u(r) = u(R),
\end{equation}
\begin{equation}
\mathop {\lim }\limits_{r \to R_1 } u(r) = \mathop {\lim }\limits_{R \to 0} u(R) = C \ne 0,
\end{equation}
 \begin{equation}
\mathop {\lim }\limits_{r \to R_1 } \frac{{R^2 }}{{r^2 }}\frac{{\partial u}}{{\partial r}} = \mathop {\lim }\limits_{R \to 0} R^2 \frac{{\partial u}}{{\partial R}} = 0,
\end{equation}                             
                                                   
In [32], by GL no scattering modeling, we proved when $j_1(kR_1) \ne 0$ , there is no solution to satisfy the the acoustic equation (8) in the inner sphere domain $R < R_1$ with boundary conditions (33) and (34) on the inner spherical surface boundary $R =R_1$  , when $j_1(kR_1) = 0$ , there is nonzero acoustic wave propagation penetrate into the inner sphere $R < R_1$ ,therefore , $0$  in the BS   to $R_1$  in the PS   and $R_2$ invariant monotone increase differentiable radial transformation can not be used to create acoustic no scattering cloak

\subsection{Comparison between GL full wave no scattering modeling method and ray tracing method for cloak}

There is no difference for the ray tracing between the acoustic and electromagnetic wave propagation  In the almost all publications. the general ray tracing method can not identify the different head terms between the radial EM equations and the acoustic equation. 
The different head terms are basic different property between the radial EM equations and the acoustic equation that is key important to study cloak. Moreover, the general ray tracing method can not identify the different essential property in theorem 3.2 in  [31] between the radial EM equations and the acoustic equation. 
\begin{equation}
\begin{array}{l}
 \frac{1}{{4\pi }}\int_0^\pi  {\int_0^{2\pi } {E^b _r (\vec r)\sin \theta d\theta d\phi  = 0} } , \\ 
 \frac{1}{{4\pi }}\int_0^\pi  {\int_0^{2\pi } {u(\vec r)\sin \theta d\theta d\phi  \ne 0} } , \\ 
 \end{array}
\end{equation}
Therefore. only ray tracing method is not complete theoretical method for studying cloak.
The GL full wave modeling and inversion[1-3] [27][31-33]is available for study the EM invisible cloak and acoustic  no scattering cloak, but the ray tracing method is not available for studying the acoustic  no scattering cloak. Moreover, because the ray tracing and wave front of the EM wave and acoustic wave propagation in GLHUA[1-3][11][27] are discontinuous, therefore, the ray tracing method is not available for studying GLHUA and GLLH[1] cloak, also the ray tracing method can not be used for studying GLHUANPII invisible cloak[33].

\section {The Radial   In The Sphere Coordinate Can Be Negative As Well As  A New  Negative Space In Which Acoustic and Electromagnetic Wave Propagation}
\subsection {We discover negative space in which the radial variabe of the sphere coordinate is negative}
Since our high school or undergraduate university, we have already been known that
the radial variable in the sphere coordinate is always positive or zero. Recent, in investigation of the wave propagation, first of the world, we discover the radial variable in the sphere coordinate can be negative. The space in which the radial variable in the sphere coordinate is positive or zero is called positive space, that is real three dimensional space we living.  The space in which the radial variable in the sphere coordinate is  negative is called negative space. The positive space is visible, but  the negative space is invisible.  In the acoustic equation in the sphere coordinate [32], radial electric wave equation and radial magnetic wave equation in the spherical coordinate [1] [31], if the radial variable $R$ is  changed to $-R$ , the equations are invariant, if the acoustic wave field solution is $u(R,\theta ,\phi )$ , the $u(-R,\theta ,\phi )$  is also solution of the homogeneous acoustic equation. Similar,  if radial electric field $E_r (R,\theta ,\phi )$  and radial magnetic wave $H_r (R,\theta ,\phi )$  are solution of the radial electric and radial magnetic equations, then 
 $E_r (-R,\theta ,\phi )$
and $H_r (-R,\theta ,\phi )$
  are also the solution of the homogeneous radial electric and radial magnetic equations  respectively, Therefore the acoustic equation and radial electric and magnetic equations and their solutions  in the sphere coordinate in the positive space with $R \ge 0$  can be analytical continuation into the negative space with $R < 0$ . That will be proved in the sub section 2 to sub section 4. We propose a GLHUANPI transformations in this paper, under  GLHUANPI transformations, the part of the negative space is translated into the other part of the positive space. The wave propagation in that part of negative space is shown in the part of the positive space. 
We propose GLHUANPI transformation and create new GLHUANPI electromagnetic permeability and dielectric in physical whole sphere and find electromagnetic and acoustic wave propagation in the whole sphere,  A new negative space in this paper is an important discoveries in the super sciences and super computational sciences in the world. We combine the negative space and positive space together to be new extended super universe 
$ - \infty  < R <  + \infty $ in sphere coordinate. The door between the positive space and negative space is $0$ .  In normal general sciences excluding religion and theology, the positive space and negative space are not connected. The door 0 between the positive space and negative space is closed. Up to now, all general sciences principles, for example, conservation of energy, are hold and satisfied in the positive space. Up to now, all scientist are working in the positive space.  We create the GLHUANPI in this paper and GLHUANPII transformation to open the door $0$ between the negative and positive space. Nothing is in  
$R <  - \infty$   or  in  $+ \infty  < R $  .The new negative space will have important theoretical and practicable applications in the sciences and super sciences. In the Pendry Cloak. the relative EM parameters are from Pendry transformation in the annular layer $R_1 < R < R_2$ , Also. Pendry put free space media with relative EM parameter $1$ in the inner sphere $0 < R < R_1$ that create Pendry invisible cloak in his paper in sciences 2006[4].In this paper, based on our novel negative space, we discovered and proved an important problem that If the Pendry relative EM  parameters  
\begin{equation}                             
 \varepsilon _r  = \mu _r  = ({{R_2 } \mathord{\left /
 {\vphantom {{R_2 } {(R_2  - R_1 )}}} \right.
 \kern-\nulldelimiterspace} {(R_2  - R_1 )}}){{(r - R_1 )^2 } \mathord{\left/
 {\vphantom {{(r - R_1 )^2 } {r^2 }}} \right.
 \kern-\nulldelimiterspace} {r^2 }}
\end{equation}
\begin{equation}
\varepsilon _\theta   = \mu _\theta   = \varepsilon _\phi   = \mu _\phi   = {{R_2 } \mathord{\left/
 {\vphantom {{R_2 } {(R_2  - R_1 )}}} \right.
 \kern-\nulldelimiterspace} {(R_2  - R_1 )}}
\end{equation}
are sit in the whole sphere  which include the a nnular  and inner sphere  , then there is nonzero EM wave in the inner sphere  , the inner sphere is not cloaked concealment.
The new idea, concept, and method have extremely significant for creating the practicable EM invisible cloak and acoustic and seismic no scattering cloak. The description arrangement of this paper is as follows: The introduction has been presented in the sub section 1; In sub s section 2, we propose and prove that Radial electric wave equation and its solution in the positive space can be analytical continuation into the negative sphere coordinate space with negative radial sphere coordinate variavle  ; The Global Local Negative to Positive space sphere GLHUANPI radial transformation GLHUANPI to map   in BS to  in PS and  in BS to  in PS is proposed in the sub section 3; In the sub section 4, we propose the electric and magnetic wave propagation through the whole sphere with radius  and with relative GLHUANPI electromagnetic parameters (3), the whole space is compressed by the GLHUANPI transformation (2) and from a negative space sphere with  and negative radius   and other positive space sphere with positive radial variable   and positive radius  ; The relationship are presented in the sub section 5; The discover are presented in the sub section 6

\subsection  { Radial Electric  wave equation in the positive and negative sphere coordinate space}

When the radial variable in the sphere coordinate space is positive or zero, the radial electric wave equation in the sphere coordinate [1][27][31] is
\begin{equation}
\frac{\partial }{{\partial R}}\frac{\partial }{{\partial R}}R^2 E_r  + \frac{1}{{\sin \theta }}\frac{\partial }{{\partial \theta }}\sin \theta \frac{{\partial E_r }}{{\partial \theta }} + \frac{1}{{\sin ^2 \theta }}\frac{{\partial ^2 E_r }}{{\partial \phi ^2 }} + k^2 R^2 E_r  = J_s 
\end{equation}
            
${\boldsymbol{Theorem \ 1:}}$ \ In the normal positive space in which the radial variable in the sphere space is positive, if the wave solution of the  electric wave equation is $E_r (R,\theta ,\phi )$ ,then $E_r (-R,\theta ,\phi )$  is also solution of homogeneous equation (1) with $J_s  = 0$ 

Proof, For point source at $(r_s ,\theta _s ,\phi _s )$ , if  $E_r (R,\theta ,\phi )$  is solution of (1) with point source. By heave calculus calculation, we prove that the  $E_r (-R,\theta ,\phi )$   satisfies homogeneous equation (38) with  $J_s  = 0$ ,
            
${\boldsymbol{Theorem \ 2:}}$ \ The homogeneous radial electric wave equation (38) with  $J_s  = 0$,  its solution $E_r (R,\theta ,\phi )$   in positive space can be analytic continuation into negative space with negative radial $R$.

Proof: Let $E_r (R,\theta ,\phi )$    be the solution of the radial electric quation (38) with point source, by theorem 1, the $E_r ({-R},\theta ,\phi )$     is also the solution
of the radial elecric wave equation,

\begin{equation}
\begin{array}{l}
\frac{\partial }{{\partial R}}\frac{\partial }{{\partial R}}R^2 E_r ( - R,\theta ,\phi ) + \frac{1}{{\sin \theta }}\frac{\partial }{{\partial \theta }}\sin \theta \frac{{\partial E_r ( - R,\theta ,\phi )}}{{\partial \theta }} \\ 
  + \frac{1}{{\sin ^2 \theta }}\frac{{\partial ^2 E_r ( - R,\theta ,\phi )}}{{\partial \phi ^2 }} + k^2 R^2 E_r ( - R,\theta ,\phi ) = 0, \\ 
 \end{array}
\end{equation}
                                   
For $\xi  < 0$, $ -\xi  > 0$, 
Substitute $ R =  -\xi  $  into the (39) , the homogeneou radial electric equation (38) with  $J_s  = 0$   and its solution that becomes
         
 \begin{equation}  
\begin{array}{l}
 \frac{\partial }{{\partial ( - \xi )}}\frac{\partial }{{\partial ( - \xi )}}R^2 E_r ( - ( - \xi ),\theta ,\phi ) + \frac{1}{{\sin \theta }}\frac{\partial }{{\partial \theta }}\sin \theta \frac{{\partial E_r ( - ( - \xi ),\theta ,\phi )}}{{\partial \theta }} \\ 
  + \frac{1}{{\sin ^2 \theta }}\frac{{\partial ^2 E_r ( - ( - \xi ),\theta ,\phi )}}{{\partial \phi ^2 }} + k^2 ( - \xi )^2 E_r ( - ( - \xi ),\theta ,\phi ) = 0, \\ 
 \end{array}
\end{equation}     
\begin{equation}             
\begin{array}{l}
 \frac{\partial }{{\partial \xi }}\frac{\partial }{{\partial \xi }}R^2 E_r (\xi ,\theta ,\phi ) + \frac{1}{{\sin \theta }}\frac{\partial }{{\partial \theta }}\sin \theta \frac{{\partial E_r (\xi ,\theta ,\phi )}}{{\partial \theta }} \\ 
  + \frac{1}{{\sin ^2 \theta }}\frac{{\partial ^2 E_r (\xi ,\theta ,\phi )}}{{\partial \phi ^2 }} + k^2 (\xi )^2 E_r (\xi ,\theta ,\phi ) = 0, \\ 
 \end{array}
\end{equation}
By using $R = \xi$, we obtain

\begin{equation}
\begin{array}{l}
 \frac{\partial }{{\partial R}}\frac{\partial }{{\partial R}}R^2 E_r (R,\theta ,\phi ) + \frac{1}{{\sin \theta }}\frac{\partial }{{\partial \theta }}\sin \theta \frac{{\partial E_r (R,\theta ,\phi )}}{{\partial \theta }} \\ 
  + \frac{1}{{\sin ^2 \theta }}\frac{{\partial ^2 E_r (R,\theta ,\phi )}}{{\partial \phi ^2 }} + k^2 (R)^2 E_r (R,\theta ,\phi ) = 0, \\ 
 \end{array}
\end{equation}

Therefore,  homogeneous radial electric wave equation (38) with  $J_s  = 0$  and its solution $E_r (R,\theta ,\phi )$   in positive space can be analytic continuation into negative space with negative radial $R$ in equation (42).
The theorem 2 is proved.

\subsection {The Global Local Negative to Positive space sphere GLHUANPI radial transformation to map $R = -h$ to  $r = 0$ and $R = R_2$ to $r = R_2$}

We can present the acoustic and electromagnetic wave propagation in the 3D positive space,   that is conventional real 3D wave propagation. However, we can not image the acoustic and electromagnetic wave propagation in the 3D negative space . In normal general sciences, the door $0$ between the positive space and negative space is closed. Fortunately, there is GLHUANPI radial transformations we proposed that can open the door and translate some part of the negative space onto the other part of the positive space. By the transformation, we can present the wave propagation in the positive and negative space together in the positive space, We present the electromagnetic wave propagation in the positive and negative space in the next sections of this paper,

\subsubsection {The linear sphere radial  GLHUANPI radial transformation to map $R = -h$ to  $r = 0$ and $R = R_2$ to $r = R_2$}
The linear sphere radial  GLHUANPI radial transformation to map $R = -h$ to  $r = 0$ and $R = R_2$ to $r = R_2$ that is
\begin{equation}
r = \frac{{R_2 }}{{R_2  + h}}(R + h),
\end{equation}
  \subsubsection {GLHUANPI electromagnetic relative parameters in the whole sphere $r\le R_2$}
By the linear sphere radial  GLHUANPI radial transformation to map $R =  - h$ to  
$ r = 0$ and $R = R_2$ to $r = R_2$ in (43),GLHUANPI electromagnetic relative parameters in the whole sphere $r \le  R_2$ are created,
\begin{equation} 
       \begin{array}{l}
 \varepsilon _\theta   = \varepsilon _\phi   = \mu _\theta   = \mu _\phi   = \frac{{R_2  + h}}{{R_2 }}, \\ 
 \varepsilon _r  = \mu _r  = \frac{{(r(R_2  + h) - R_2 h)^2 }}{{R_2^2 r^2 }}\frac{{R_2 }}{{R_2  + h}}, \\ 
 \end{array}
\end{equation} 
 \subsubsection{When $0  <  R_1  <  R_2$  and $h = \frac{{R_1 R_2 }}{{R_2  - R_1 }}$, the GLHUANPI transformation in (43) does become Pendry transformation in [4]  }
For  $0 < R_1 < R_2$   substitute 
\begin{equation}
h = {{R_1 R_2 } \mathord{\left/
 {\vphantom {{R_1 R_2 } {(R_2 }}} \right.
 \kern-\nulldelimiterspace} {(R_2 }} - R_1 ),
\end{equation}
  into (43),

\begin{equation}
r = \frac{{R_2 }}{{R_2  + h}}(R + h) = \frac{{R_2 }}{{R_2  + \frac{{R_1 R_2 }}{{R_2  - R_1 }}}}(R + \frac{{R_1 R_2 }}{{R_2  - R_1 }}),
\end{equation}

\begin{equation}
r = \frac{{R_2 }}{{R_2  + h}}(R + h) = \frac{{R_2  - R_1 }}{{R_2 }}(R + \frac{{R_1 R_2 }}{{R_2  - R_1 }}),
\end{equation}
\begin{equation}
\begin{array}{l}
 r = \frac{{R_2 }}{{R_2  + h}}(R + h) \\ 
  = \frac{{R_2  - R_1 }}{{R_2 }}(R + \frac{{R_1 R_2 }}{{R_2  - R_1 }}) \\ 
  = R_1  + \frac{{R_2  - R_1 }}{{R_2 }}R, \\ 
 \end{array}
\end{equation}
the Pendry transformation [4] is obtained.

\subsubsection{When $0 < R_1 < R_2$  and $h = \frac{{R_1 R_2 }}{{R_2  - R_1 }}$, The GLHUANPI electromagnetic relative permeability and dielectric parameters (44) become Pendry parameters in [4]}                                                                            
Substitute (45) into (44), Pendry EM parameters are obtained in the whole sphere 

\begin{equation}
\begin{array}{l}
 \varepsilon _r  = \mu _r  = \frac{{R_2 }}{{R_2  - R_1 }}\frac{{(r - R_1 )^2 }}{{r^2 }}, \\ 
 \varepsilon _\theta   = \mu _\theta   = \varepsilon _\phi   = \mu _\phi   = \frac{{R_2 }}{{R_2  - R_1 }}, \\ 
 \end{array}
\end{equation}

 \subsection { The electric and magnetic wave propagation through the whole sphere with radius $R_2$  and with relative GLHUANPI electromagnetic parameters (44) }

\subsubsection {The electric  wave propagation through the whole sphere with radius  and with relative GLHUANPI electromagnetic parameters (44)} 

By the GLHUANPI radial transformation (43), the basic  negative sphere
\begin{equation}
 - h \le R < 0,
\end{equation}                                                                                                     
in the negative space is translated onto the physical positive sphere  
\begin{equation}       
0 \le r < \frac{{R_2 h}}{{R_2  + h}},
\end{equation}
in the positive space, other basic positive sphere   $0  \le R < R_2 $ in the positive space is translated onto the physical annular layer $ \frac{{R_2 h}}{{R_2  + h}} \le r \le R_2$ , the outside of the sphere in the physical positive space is not changed.  The electric wave propagation in the basic  negative sphere $- h \le R < 0,$  in the negative space is compressed and presented in the physical positive sphere $0 \le r < \frac{{R_2 h}}{{R_2  + h}},$ in the positive space; electric wave propagation in the basic  positive sphere $0 \le R < R_2 $  in the positive space is compressed and presented in the physical positive annular layer $ \frac{{R_2 h}}{{R_2  + h}} \le R \le R_2$   in the positive space; electric wave propagation is not changed in the outside of the sphere $r \ge R_2$ , Using theorem 1 and GL electromagnetic modeling simulation, the image of the electric wave propagation in the positive space and negative sphere $-h \le R \le 0$   , i.e, is that the physical electric wave propagation through whole sphere $0 \le r <R_2$   with relative GLHUANPI electromagnetic parameters (44) is present in the Figure 1.

\subsubsection { The magnetic wave propagation through the whole sphere with radius $R_2$  and with relative GLHUANPI electromagnetic parameters (44)}

Using GL electromagnetic modeling simulation, the magnetic wave propagation through the whole sphere with radius  $R_2$   and with relative GLHUANPI electromagnetic parameters (44)  is presented in Figure 2.

\subsection { Relationship of the GLHUANPI transformation and Pendry Transformation}

\subsubsection {In the GLHUANPI transformation (43), we chose parameter $0 < h < \infty $}

\subsubsection { For $0 < R_1 < R_2$  and $h = \frac{{R_1 R_2 }}{{R_2  - R_1 }}$  , the GLHUANPI transformation in (43) does become Pendry transformation (48) in [4]} 

 The GLHUANPI electromagnetic relative permeability and dielectric parameters (44) does  become Pendry parameters (49) in [4]. However, Pendry only did sit their relative parameters in the annular layer  $R_1 < r < R_2  $, and let free space in the inner sphere $0 < r < R_1$    and obtain impracticable and  fundamental difficult EM invisible cloak with infinite speed and exceeding light speed propagation in the cloak. 

\subsubsection { We discover the negative space and in which the electromagnetic and acoustic wave propagation. We propose the  GLHUANPI transformation (43) with  $0 < h < \infty $  and  discover the GLHUANPI electromagnetic parameters (44)  in the whole sphere.} 

\subsubsection {By GL no scattering modeling simulations, the electromagnetic propagation in the whole sphere with the GLHUANPI electromagnetic parameters (44)  are presented in the Figure 19.  The Figure 19 shows the radial electric wave propagation through the whole sphere $0 \le r \le R_2$ , The inner spher is not cloaked concealment} 

\subsubsection { If Pendry EM parameters [4] in the annular layer $R_1 \le r \le R_2$ are sit into
the whole sphere $0 \le r \le R_2$  which include the annular layer $R_1 \le r \le R_2$  and inner sphere $0 \le r \le R_1$. What happen for the electromagnetic wave propagation through the whole sphere $0 \le r \le R_2$    ? There is electromagnetic wave propagation in the inner sphere $0 \le r \le R_1$ ?}

We discover and prove the following theorem

${\boldsymbol{Theorem \ 3:}}$ \ When the GLHUANPI electromagnetic parameters (44)  are sit in the whole sphere  $0 \le r \le R_2$ , there is no any sub inner sphere can be cloaked. For example, if the Pendry parameters in [4] are put in the whole sphere $0 \le r \le R_2$ , then inner sphere $0 \le r \le R_1$.    can not be cloaked concealment.

From the  theorem 3 , we prove that

${\boldsymbol{Theorem \ 4:}}$ \ If Pendry EM parameters are sit in the whole sphere, $0 \le r \le R_2$   which include the annular layer  $R_1 \le r \le R_2$   and inner sphere $0 \le r \le R_1$ , then there is the
electromagnetic wave propagation through whole sphere $0 \le r \le R_2$     , there is nonzero
electromagnetic wave propagation in the inner sphere $0 \le r \le R_1$ , the inner sphere is not
cloaked concealment. 

The radial electric wave propagation through whole sphere and there exist nonzero electro wave in the inner shere that is shown in Figure 19. the radial electric wave 
ecited by Delata point electric souce at $r_s > > R_2$    Figure 20 presents that Pendry EM parameters are sit in the whole sphere, $0 \le r \le R_2$     which include the annular layer $R_1 \le r \le R_2$    and inner sphere $0 \le r \le R_1$ ,

\subsection{Discover }

\subsubsection {In this paper, we discovered the negative space in which the radial variable in the 
 sphere coordinate is negative.}

\subsubsection {We discovered proved that the acoustic wave and electromagnetic wave equations and their wave solution in the positive space can be analytical continuation into the negative space.}

${\boldsymbol{Theorem \ 5:}}$ \ The homogeneous electromagnetic wave equation ,  its solution $\vec E (R,\theta ,\phi )$    and $ \vec H (R,\theta ,\phi )$    in positive space can be analytic continuation into negative space with negative radial $R$.

${\boldsymbol{Theorem \ 6:}}$ \ In the normal positive space in which the radial variable in the sphere space is positive, if the wave solution of the acoustic wave equation is $u (R,\theta ,\phi )$ ,then $u (-R,\theta ,\phi )$  is also solution of homogeneous equation (1) with $J_s  = 0$ 

${\boldsymbol{Theorem \ 7:}}$ \ The homogeneous  acoustic wave equation (38) with source zero,  its solution $u (R,\theta ,\phi )$   in positive space can be analytic continuation into negative space with negative radial $R$.

\subsubsection {The new negative space will have important theoretical and practicable applications in the sciences and super sciences and super computational sciences, The new idea, concept, and method have extremely significant for creating the practicable EM invisible cloak and acoustic and seismic no scattering cloak. The copyright, patents, and all rights are reserved by authors in GL Geophysical Laboratory, USA}

\section{Discussion and Conclusion}

We have discovered GLHUA exact analytic EM full wave solution (25) and (33 of 3D MAXWELL equation in GLHUA double layer invisible cloak (21) and (23) . Our analytical EM wave in GLHUA cloak is undisputed evidence and rigorous proof to prove that the GLHUA double layer cloak is practicable invisible cloak without exceed light speed propagation. Our analytical EM wave in GLHUA cloak is undisputed evidence and rigorous proof to prove GLHUA double layer invisible cloak and their theoretical proof in  are right. 

GL simulation shows that there are novel GLHUA mirage bridge which is generated by the
GLHUA outer layer invisible cloak materials in (21) and (23). The GLHUA mirage bridge make the GLHUA analytical EM wave propagation without exceeding light speed. The GLHUA mirage bridge  is obvious for $R_1=0.5R_2$ or $R_1=0.66R_2$,
in our new paper with arxiv submitted ID submit1928169 submitted on June 21,2017[33]. For $R_> 0.8R_2$, the GLHUA mirage bridge wave become to the arc  GLHUA mirage wave on the inner spherical layer $r=R_1$, the obvious mirage bridge is disappear in nxet paper.  
In the figures 2-14 in this paper,$R_1=0.66R_2$, we did simulation for magnetic wave propagation in GLHUA outer layer cloak material in (1) which is different from (21) and (23) in [33], the mirage bridge are dispear.  The GLHUA mirage bridge generated by GLHUA outer layer cloak material (1) is obvious for $R_1=0.5R_2$.
The analytical GLHUA EM wave show that in the any finite time the EM wave in GLHUA outer layer cloak can not be arrive to the inner spherical boundary,$r = R_1 $, that making the EM wave can not penetrate into the sphere $r < R_1 $. Therefore, the analytical GLHUA EM wave proved that GLHUA outer layer cloak is invisible cloak with sphere concealment $r < R_1 $. 
In generalized regularizing sense, when $r$ going to $R_1$, our GLHUA analytical EM wave field (25) and (33) are going to zero.

The strange double layer cloak phenomena in Figure 1 in 2000 are image of the nonzero solution of GILD no scattering inversion which cloaked the object target of GILD scattering inversion. After 16 year hard works, GLHUA invisible EM double layer cloak in this paper presented a novel model and theory to explain the observation of the double cloak phenomena in 2000 [3]. The current physical experiment is scattering experiment, so the conventional physical experiment can not be used for studying invisible cloak. Pendry cloak [4] and ULF cloak [10] are created by the mathematical coordinate transform. There is no experiment to support the invisible cloak. There is no any complete invisible cloak that is discovered by conventional physical experiment. There is no any complete invisible cloak material is discovered by exploration in natural world. The material of Pendry cloak can not be found in natural world. The phase velocity is exceeding light speed and tends infinite, and reciprocal principle is not satisfied that is three fundamental difficulties in Pendry cloak. The three fundamental difficulties can not be solved by the optical coordinate transform method.  Global and Local field GL modeling and no scattering inversion is powerful method to create practicable GLHUA double layer cloak and play key important role to prove GLHUA double layer cloak to be invisible cloak with concealment that overcome the above three fundamental difficulties in Pendty cloak. HUA is from name of first author, Hua also means that H-Up-(Alpha)=$h^\alpha  $ where $h=r - R_1$. So, we call GLHUA cloak. The original material in GLHUA cloak in (1) and (2) can be found in the natural science. The practicable fabrication of GLHUA cloak need to be investigated in the future. GL no scattering inversion method for cloak is to find a material in local device that making zero scattering. No scattering is absolute zero scattering that is different from small scattering. Complete invisible cloak is different from the partial or approximate invisible cloak. 
GL simulation shows that the relative GLHUA parameter is not unique. Nonzero solution set of the no scattering inversion is complicated infinite class presented (1) and (2) (in (1) and (12) in [1]). We rigorous proved GLHUA double layer cloak with $\alpha  = 1$ that is invisible cloak with concealment. The detailed proof is presented in [1] and [29]. For $0 < \alpha _0  < \alpha  < \alpha _1  < 2$, we can similarly prove that GLHUA double layer cloak with materials in (1) and (2) ((1) and (12) in [1] ) is invisible cloak with concealment. 
The slowness of the material in GLHUA cloak is large than or equal to 1, why does the light ray through GLHUA cloak can arrive outside of the cloak at same time as with in free space. That is because the EM wave front and ray in GLHUA outer layer cloak are discontinuous at the inner boundary,  $r=R_1$ , of the outer layer cloak. The theorem 4.1-4.4 in section 4 in [1] prove that, in GLHUA outer layer cloak, EM wave propagation is going to zero in the inner boundary  $r=R_1$ , that is key important property to prove GLHUA outer cloak is invisible cloak with concealment. The key property making that phase velocity of GLHUA cloak is no exceeding light speed and tends to zero. The key property makes that the EM wave front in GLHUA cloak is discontinuous and splitting into upper front and lower front on the inner spherical surface boundary $r=R_1$  which are shown in Figure 2 to Figure 14. GLHUA double layer cloak material with relative parameter (1) and (2) ((1) and (12) in [1]) presents absorption and emission effects. The incoming EM full wave field is absorbed to zero at the inner boundary by GLHUA material that making incoming wave ray to be terminated and discontinuous at the inner boundary. The discontinuous upper and lower front, climbing up and sliding down along and pass inner spherical surface $r=R_1$ , then wave front is recovered. The ray is re born by GLHUA material emission effect. That making wave propagation scattering is zero by cloak, incident wave excited in outside of cloak is not to be disturbed by cloak, that making GLHUA cloak is invisible that are rigorously proved in
[1] and [29]. Because GLHUA cloak material absorbed the EM wave to zero at the inner spherical surface $r=R_1$ , that making the EM wave can not penetrate to concealment. The Eikonal equation (144) and transport equation  in [1] by [24-25] for EM wave field in anisotropic cloak media must be jointly solved simultaneous and connection equation in section 8 that is different from traditional Eikonal equation and geometry ray equation. Therefore, the traditional Eikonal equation and geometry ray method with travel time only and without intensity can not be used to study and explain GLHUA cloak [16]. In Pendry cloak, wave propagation front and ray are continuous and curved by linear coordinate transform, but the wave front and ray in Pendry cloak are not orthogonal intercross each other. In Pendry cloak, the inner annular boundary,  $r=R_1$  is included in one wave front that caused infinite phase velocity.
  In paper by Pendry et al in Science in 2006 [4], Pendry did state that nor can any radiation get out. that is totally wrong. We in [8][15] and Zhang and Chen in [7] proved that any radiation excited in Pendry cloak concealment will be propagation go to out. The reciprocal principle is not satisfied for Pendry cloak and is not satisfied for any single layer cloak. For overcoming the difficulty, we proposed inner layer cloak in GLHUA double layer cloak in this paper and in [9] in 2009. Up to now, only our inner cloak of the double layer cloak is discovered and proposed. The EM wave excited in the concealment $r < R_0$ can not be disturbed by cloak. The wave propagation can be not arrive $r=R_1$ and can be not go out of inner layer cloak. The wave propagation in GLHUA inner layer cloak is shown in Figure 17. The double layer 
$[R_0, R_1]$ and $[R_1, R_2]$  in GLHUA double layer cloak can be splitting into the inner layer $[R_0, R_1]$ with parameter (2) ((12) in [1]) and the outer layer $[R_2, R_3]$  with parameter
\begin{equation}
\begin{array}{l}
 \varepsilon _r  = 1, \\ 
 \varepsilon _\theta   = \varepsilon _\phi   = \frac{1}{2}\left( {\left( {\frac{{r - R_2 }}{{R_3  - R_2 }}} \right)^\alpha   + \left( {\frac{{R_3  - R_2 }}{{r - R_2 }}} \right)^\alpha  } \right), \\ 
 \end{array}     
\end{equation}
in (1) ((1)in [1]), using $R_2$ to replace $R_1$ and $R_3$ to replace $R_2$, the middle layer 
$[R_1, R_2]$ can be free space or water to adjust temperature in practice.
GLHUA cloak and GLLH cloak published in paper arXiv1005.3999 are different class. Some numerical dispersion in GL method simulation of GLLH cloak in the paper arXiv1005.3999 [11] has been improved.  
Chen at al proved Pendry Cloak is invisible by Mei transform method [28], which is only available for constant relative angular parameter. 
Using GL no scattering modeling and inversion, we easy to prove that in Pendry cloak, 

\begin{equation}
\begin{array}{l}
 \mathop {\lim }\limits_{r \to R_1 } E\left( {\vec r} \right) = r^2 \varepsilon _r E_r  = 0, \\ 
 \mathop {\lim }\limits_{r \to R_1 } H\left( {\vec r} \right) = r^2 \mu _r H_r \left( {\vec r} \right) = 0, \\ 
 \end{array} 
\end{equation}

\begin{equation}
\begin{array}{l}
 \mathop {\lim }\limits_{r \to R_1 } \frac{1}{{\varepsilon _\theta  }}\frac{\partial }{{\partial r}}E\left( {\vec r} \right) =  \\ 
  = \mathop {\lim }\limits_{r \to R_1 } \frac{1}{{\varepsilon _\theta  }}\frac{\partial }{{\partial r}}\left( {r^2 \varepsilon _r E_r \left( {\vec r} \right)} \right) = 0 \\ 
 \end{array}             
\end{equation}
\begin{equation}
\begin{array}{l}
 \mathop {\lim }\limits_{r \to R_1 } \frac{1}{{\mu _\theta  }}\frac{\partial }{{\partial r}}H\left( {\vec r} \right) =  \\ 
  = \mathop {\lim }\limits_{r \to R_1 } \frac{1}{{\mu _\theta  }}\frac{\partial }{{\partial r}}\left( {r^2 \mu _r H_r \left( {\vec r} \right)} \right) = 0, \\ 
 \end{array}                 
\end{equation}
By theorem 51-5.6 in section 5 of the paper [1], we proved that Pendry Cloak is invisible cloaking for incident wave excited in outside of the cloak. However, any incident wave excited in the concealment will propagate to outside of the cloak, or that causes the system has no solution The reciprocal principle is not satisfied in Pendry cloak. Using GL modeling, we did many simulations for Pendry cloak. By comparison , GL simulation imaging show that the phase velocity of the EM wave propagation through GLHUA cloak is less than light speed and tends to zero at boundary  $r=R_1$. However, the phase velocity of EM wave propagation through Pendry cloak is with exceeding light speed and tends to infinity at boundary  $r=R_1$. Our GLHUA double layer cloak does overcome the three fundamental difficulties of Pendry Cloak.
Using the above proof method we proved that any annular layer cloaking by optical transform is invisible cloaking with infinite phase velocity and excceding light speed for incident wave excited in outside of the cloak. But, any incident wave excited in the concealment will propagate to outside of the cloak. However, the $0$ to $R_1$ spherical radial transformation can not be used for acoustic cloak.[32]
In our paper [5-6][11-12][15], we proposed GL no scattering modeling and inversion to create practicable GLHUA and GLLH two class of the invisible cloak. Our GLHUA cloak by GL no scattering modeling and inversion is different from GLLH cloak in 2010. In 1971, we used no scattering inversion idea to construct a novel 3D 20 nodes high accurate curve element and developed 3D finite element method first in China and discovered super convergence of 3D iso parameter finite element first in the world [26][34]. GILD and GL scattering and no scattering modeling and inversion idea and method play very important role for research works in invisible cloak. When Pendry cloak in 2006, we had been working in no scattering modeling and inversion for 6 years since 2000. Our idea and method are different from Pendry and other cloak research.
Any annular layer cloaking by optical transform will make the phase velocity exceeding light speed and tends to infinity in $r=R_1$. 
Because there is no experiment to study the complete invisible cloak, therefore the full wave theoretical proof and full wave computational simulation are necessary to verify the invisible clock. GLHUA invisible cloak and GLLH invisible cloak have been verified by the full wave theoretical proof and full wave simulation using GL full wave no scattering modeling and inversion in [1]. 
GLHUA double cloak and mirage in [13] show that GL no scattering modeling and inversion is powerful method to make practicable invisible cloak and to investigate new invisible natural science Inventor Easton LaChappelles mind control robot hands is beginning to investigate visible thinking science and visible social Science. In current general science, visible natural science and invisible thinking science and social science are main object. Now invisible natural science and invisible thinking science and visible social science will be main object for a novel super science. Our GLHUA practicable double cloak and Easton LaChappelles mind control robot hands show that the new novel super science is being born.
\begin{acknowledgments}
We wish to acknowledge the support of the GL Geophysical Laboratory and thank the GLGEO Laboratory to approve the paper
publication. Authors thank to Professor P. D. Lax for his concern and encouragements  Authors thank to Dr. Michael Oristaglio and Professor You Zhong Guo for their encouragments
\end{acknowledgments}


\end{document}